\documentclass[12pt]{article}
\usepackage[utf8]{inputenc}
\usepackage[titletoc,title]{appendix}
\usepackage{amsmath,amsfonts,amssymb,mathtools}
\usepackage{graphicx}
\usepackage{algorithm,algpseudocode}
\usepackage{array}
\usepackage{nameref}
\usepackage[table]{xcolor}
\newcommand*\hy{\color{black}} 
\newcolumntype{x}[1]{%
>{\centering\hspace{0pt}}p{#1}}
\newcommand{\tn}{\tabularnewline}
\graphicspath{ {./images/} }

\usepackage{shortcut}
\usepackage{natbib}

\usepackage{titlesec}
\titleformat{\section}
  {\normalfont\fontsize{12}{15}\bfseries}{\thesection}{1em}{}
  
\titleformat{\subsection}
  {\normalfont\fontsize{12}{15}\bfseries\itshape}{\thesubsection}{1em}{}
  
\titleformat{\subsubsection}
  {\normalfont\fontsize{12}{15}\itshape}{\thesubsubsection}{1em}{}

\newcommand{\blind}{1}

\begin{document}
\def\spacingset#1{\renewcommand{\baselinestretch}%
{#1}\small\normalsize} \spacingset{1}


\if1\blind
{
  \title{\bf A Process of Dependent Quantile Pyramids}
  \author{Hyoin An\thanks{Correspondence: Hyoin An (an.355@osu.edu), Department of Statistics, The Ohio State University}  \hspace{.05em} and  Steven N. MacEachern \\
    Department of Statistics, The Ohio State University}
  \maketitle
} \fi

\bigskip
\begin{abstract}

Despite the practicality of quantile regression (QR), simultaneous estimation of multiple QR curves continues to be challenging.
We address this problem by proposing a Bayesian nonparametric framework that generalizes the quantile pyramid by replacing each scalar variate in the quantile pyramid with a stochastic process on a covariate space. 
We propose a novel approach to show the existence of a quantile pyramid for all quantiles. 
The process of dependent quantile pyramids allows for nonlinear QR and automatically ensures non-crossing of QR curves on the covariate space.
Simulation studies document the performance and robustness of our approach. An application to cyclone intensity data is presented.

\end{abstract}

\noindent%
\vfill

\spacingset{1.9} 
\section{Introduction}

\subsection{Quantile Regression}
Quantile regression (QR) has drawn increased attention as an alternative to mean regression. 
QR was motivated by the realization that extreme quantiles often have a different relationship with covariates than do the centers of the response distributions.  QR can target quantiles in the tail of the distribution and is more robust to outliers than is mean regression.  The advantages of QR can be substantial and have led to its use in many areas, including econometrics, finance, medicine, and climatology.  

The seminal work of \cite{koenker1978regression} extended median regression, which dates back at least as far as \cite{edgeworth1888mathematical}, to QR by allowing asymmetry in the objective function defining the regression.  That is, when one is interested in the $\tau^{th}$ quantile ($0<\tau<1$) for the response $y_i$ and the covariate $x_i \in \mathbb{R}^{p+1}$, $i = 1, \ldots, n$, and assuming independent responses, the estimated QR surface is $x_i^T b^*$, where
$b^* = \arg\min_{b \in \mathbb{R}^{P+1}} \sum_{i=1}^n \rho_{\tau}(Y_i - x_i^T b),$ 
and $\rho_{\tau}(u) = u(\tau - 1_{(u<0)})$ is the (asymmetric) check loss function.  This method is implemented in the R package `quantreg' \citep{koenker_2005} and has led to a wide range of developments.  A recent overview of the area is provided in \cite{koenker2017quantile}.

The Bayesian counterpart of quantile regression was introduced by \cite{yu2001bayesian} who used the asymmetric Laplace distribution (ALD) for the sampling density of $Y \vert x$ as a device to focus on the QR for the $\tau^{th}$ quantile.  The ALD has density $f(u) = \tau(1-\tau)\exp (-\rho_{\tau}(u))$ which can be seen as a scaled and exponentiated check loss function.  This substitution of a loss function for the log-density is an early example of the generalized Bayes technology developed in \cite{bissiri2016general}.  \cite{kozumi2011gibbs} made use of the reparametrization of the ALD illustrated in \cite{kotz2001laplace} and \cite{tsionas2003bayesian} to create an efficient Gibbs sampling algorithm.  The method is implemented in the R package `bayesQR' \citep{benoit2017bayesqr}.  Many authors have followed the approach of \cite{yu2001bayesian}, 
describing properties of the method.  See, for example, \cite{geraci2007quantile, reich2010flexible, waldmann2013bayesian, lum2012spatial}.  Some have appealed to a semi- or nonparametric approach. 
\cite{kottas2009bayesian}, for instance, proposed two approaches to model the error distribution nonparametrically in QR, using a Dirichlet process (DP) mixture of uniform densities and a dependent DP mixture of normal densities.  \cite{chen2009automatic} developed a QR function in a nonparametric fashion using piecewise polynomials.

\subsection{Crossing quantiles}
When more than one quantile level is considered, however, fitting a QR curve for each level by itself does not correspond to an encompassing model, may not respect the monotonicity of the quantile function, and can result in crossing quantiles.  Researchers have suggested various approaches to handle this issue.  \cite{rodrigues2017regression} constructed a likelihood inspired by Yu \& Moyeed's approach, while ensuring monotonocity of quantiles with an additional adjustment step.
Semi- or non-parametric Bayesian approaches to simultaneous QR include \cite{taddy2010bayesian} who suggested an approach to estimate the entire joint density of $(x, y)$ and then extract the QR from this density. This ensures monotonicity of quantiles since the inference of quantiles is based on a single density.  \cite{reich2011bayesian} and \cite{reich2012spatiotemporal} model the entire quantile process using Bernstein polynomial basis functions in spatial and spatiotemporal settings. In both papers, the prior is specified to satisfy the monotonicity constraint on the quantile function.  \cite{kadane2012simultaneous} developed a characterization of the quantile function 
that induces monotonicity in the joint estimation of linear QR models for a univariate covariate.
\cite{yang2017joint} extended this to any bounded covariate space in $\mathbb{R}^P$ via reparameterization.  \cite{chen2021joint} generalized this to {\hy incorporate spatial dependence}.

\cite{hjort2009quantile} proposed the \textit{quantile pyramid} (QP) for nonparametric inference for a single distribution and briefly mentioned a possible extension to QR. 
Most similar to our approach, \cite{rodrigues2019pyramid} used the QP for QR. In their work, independent QPs are used to specify the prior distribution for quantiles at $(p+1)$ pivotal locations in a bounded $p$ dimensional covariate space.  For each quantile, a linear QR is then constructed as the hyperplane passing through the specified quantile at each of the pivotal locations.
\cite{rodrigues2019simultaneous} adapted this idea to a spline regression setting.


\subsection{Our contribution}
We generalize the construction of \cite{hjort2009quantile} by incorporating dependence in the QPs across the covariate space and by allowing for non-binary splits in the pyramids.  Our approach allows direct and flexible modeling of the quantiles over covariate spaces and, by construction, naturally respects the monotonicity of QR curves. 

Our contribution is twofold: (1) a novel approach to show the existence of a single QP, and (2) extension of the QP from a model for a single distribution to a model for a collection of distributions that vary with the covariate. The first point is a stepping stone to generalize the idea of QP.  With an eye to the second point, it also involves expansion of the mathematical framework to move from a single QP to a process of QPs, with greater attention to mappings between the interval [0,1] and the real line. 

The rest of this article is organized as follows. In Section~\ref{sec:DQP}, we introduce the idea of a process of dependent quantile pyramids (DQPs) and a canonical construction of the model. Section \ref{sec:theory} provides theoretical results. Section~\ref{sec:posteriorinference} describes prior specification and posterior inference. Simulation studies appear in Section~\ref{sec:simulation} and application to real data appears in Section~\ref{sec:cyclone}. Section~\ref{sec:discussion} presents discussion and directions for future work.

\section{A Process of Dependent Quantile Pyramids}
\label{sec:DQP}

In this section, we briefly recap the QP of \cite{hjort2009quantile} and introduce a DQP. The following remark comes from \cite{parzen2004quantile}. 

\begin{remark}
For a random variable $Y$ whose distribution function is $F(\cdot)$, its quantile function is defined as a left-continuous function $Q(\tau) \equiv F^{-1}(\tau) = \inf\{y: F(y) \ge \tau\}$ that satisfies 
$F(y) \ge \tau$ if and only if $y \ge Q(\tau)$ for $0 < \tau < 1$.
\end{remark}
In other words, if we define the quantile function, there exists a random variable with the corresponding distribution function. This fact is useful to understand how constructing quantile functions can lead to a distribution over distribution functions.

\subsection{Binary Quantile Pyramid}
\label{sec:quantilepyramid}

\cite{hjort2009quantile} created the QP, a Bayesian nonparametric model that focuses on the quantiles of a distribution.  The QP provides a distribution over distribution functions, and so it is suited to use as a prior distribution for an unknown distribution function. The quantile function, $Q(\cdot)$, on the unit interval $[0, 1]$, is at the heart of the QP.  $Q(0) \equiv 0$ and $Q(1) \equiv 1$.  

The pyramid is built in levels for dyadic quantile levels, as a binary tree.  The $0^{th}$ level is the unit interval $[0, 1]$. At the first level, the median of the unit interval is drawn from some density, which divides the interval into two subintervals.  Intervals are recursively split into smaller subintervals, doubling the number of subintervals with each new level.  Thus, at level $m$, we have specified the $2^m-1$ quantiles, $Q^m(i/2^m)$, $i = 1, 2, \ldots, 2^m-1$.  The quantiles $Q^m(j/2^m) \equiv Q^{m-1}(j/2^m)$, $j = 2, 4, \ldots, 2^m-2$, are inherited from level $m-1$.  The new quantiles at the $m^{th}$ level can be expressed, for $j = 1, 3, \ldots, 2^m-1$, as
\begin{equation}
\label{eq:qpconstruction}
Q^m(j/2^m) = Q^{m-1}((j-1)/2^m)(1-V_{m, j}) + Q^{m-1}((j+1)/2^m)V_{m, j},
\end{equation}
where $V_{m,j}$, $j = 1, 3, \ldots, 2^m-1$, are a set of mutually independent random variables with support $[0,1]$.  {\hy Figure \ref{fig:DQPidea} (a) contains a visualization of this idea with three quantiles.} For $\tau \in (0,1)$, less the specified quantile levels, $Q(\tau)$ is filled in by linear interpolation. 

There is scope for a wide variety of choices for the distribution of the conditional medians of the subintervals (the $V_{m,j}$). 
If the $V_{m,j}$ are assumed to have mean $1/2$, for example, then $Q^m(\tau)$ forms a martingale sequence and has a limit almost surely by Doob's martingale convergence theorem.  Moreover, if $V_{m,j}$ are chosen so that $\max_{j \le 2^m}\{Q^m(j/2^m) - Q^m((j-1)/2^m) \overset{p}{\to} 0$, \cite{hjort2009quantile} showed that there exists a continuous limiting quantile process to which $Q^m$ converges. 

{\hy While \cite{hjort2009quantile} primarily focused on quantiles at the dyadic levels, we take inspiration from the \textit{oblique pyramid} construction method introduced by \cite{rodrigues2019pyramid}.  In the oblique pyramid, quantile levels in a binary pyramid are not limited to dyadics.
}

\subsection{General Quantile Pyramid}
\label{sec:generalQP}

{\hy
\cite{rodrigues2019pyramid} focus on finite quantile pyramids, where a pyramid has only $M$ levels.  They begin with scenarios where $2^{m-1}$ quantiles are generated at the $m^{th}$ level of a pyramid, ensuring that each subinterval contains one specified quantile. This constraint leads to exactly $2^M-1$ quantiles in an $M$-level pyramid.  They then proceed to the oblique pyramid, a pyramid in which some subintervals contain a quantile but others do not.  An $M$ level pyramid may be unbalanced, specifying fewer than $2^M-1$ quantiles.  

\cite{rodrigues2019pyramid} specify a rule for how the pyramid is constructed.  We provide a brief example and refer the reader to their paper for full details.  Consider the case of four quantiles, denoted as $Q(\tau_1) < \ldots < Q(\tau_4)$.  Since there are an even number of quantiles, there are two quantiles positioned in the middle. Following {\hy the rule of \cite{rodrigues2019pyramid}}, the smaller quantile serves as the middle quantile level. Thus, $Q(\tau_2)$ is specified at the first level of the pyramid.  
Moving on to the second level, we have two subintervals.  In the left subinterval, we specify $Q(\tau_1)$, while the right subinterval contains two quantiles.  The smaller quantile, $Q(\tau_3)$, is specified at this level of the pyramid.  In the third level, there are four subintervals: $(0, Q(\tau_1))$, $(Q(\tau_1), Q(\tau_2))$, $(Q(\tau_2), Q(\tau_3))$, $(Q(\tau_3), 1)$.  In this level, we specify the remaining quantile, $Q(\tau_4)$, in the last subinterval, while the other three subintervals remain empty.
}
{\hy This example is illustrated in Figure~\ref{fig:DQPidea} (b).}

{\hy
We introduce additional flexibility by permitting non-binary splits.  This yields the {\it general quantile pyramid}, and it empowers the user to fully customize the pyramid according to their preference. For instance, in the case of four quantiles, one can create a pyramid with $Q(\tau_2)$ at the first level and the remaining three quantiles at the second level, distributing one quantile to the left subinterval and two to the right subinterval. This concept is visually represented in Figure~\ref{fig:DQPidea} (c).  The move from a binary pyramid to a general pyramid requires novel notation which is introduced in the next section.  
}

\subsection{Dependent Quantile Pyramids}
\label{sec:dqp}

In the sequel, we extend the QP from a single distribution to a collection of distributions, creating {\it a process of dependent quantile pyramids} (DQPs).  To do so, we construct a QP at each value of $x$ in some index set, $\mathcal{X}$.  
We replace each scalar $V_{m,j}$ in equation (\ref{eq:qpconstruction}) with an appropriate stochastic process, say $\{V_{x,m,j}$, $x \in \mathcal{X}\}$.  This leads to a collection of QPs that may exhibit dependence across ${\cal X}$.  Formally, we construct a distribution-valued stochastic process.  For modeling purposes, the index set of the process is identified with the covariate space, as in QR.  Alternatively, the index set may be described as containing values of predictors, spatial locations, constructed features, or time, to name a few possibilities.  The important part is that, conditional on $x$, we have a model for a QP.  
We construct QPs on the unit interval $[0,1]$ following \cite{hjort2009quantile} and use a general quantile pyramid scheme of Section~\ref{sec:generalQP}.  
{\hy Figure~\ref{fig:DQPidea} (d) contains the visualization of an example of a process of DQPs.}


\subsubsection{Notation}

{\hy
We write $Q_x(\tau)$ for the $\tau$-quantile at $x$. For any $x$, $\{Q_x(\tau), \tau \in (0,1)\}$ forms a complete quantile pyramid.  For any $\tau$, $\{Q_x(\tau), x \in \cal{X}\}$ provides the $\tau$-quantile surface over the covariate space.  We refer to this as a QR curve.  The DQP is constructed sequentially.  To provide intuition behind the DQP and to facilitate proofs of its existence, we establish two distinct sets of notation.  }

{\hy The first set of notation matches the recursive construction of the DQP.  We begin with a DQP with $m-1$ levels, where the quantile surfaces have been specified for $\tau \in \mathcal{T}_{m-1}$.  That is, for $x \in {\cal X}$ and $\tau \in \mathcal{T}_{m-1}$, the function $\{ Q_{x}(\tau), x \in \cal{X}, \tau \in$ $\mathcal{T}_{m-1}\}$ has already been determined in the sequential construction.  The ordered quantile levels are $\mathcal{T}_{m-1} = \{ \tau_1^*, \ldots, \tau_T^* \}$, with $\tau_1^* < \ldots < \tau_T^*$.   At the next level, these quantile levels remain.  The corresponding quantiles define a set of bands within which the quantiles newly specified at level $m$ of the pyramid must lie (at a given $x$, the band is merely an interval).  For example, the QR curve for a quantile $\tau \in (\tau_1^*, \tau_2^*)$ that is newly specifed at level $m$ must lie in the band with lower boundary $\{Q_x(\tau_1^*), x \in \cal{X}\}$ and upper boundary $\{Q_x(\tau_2^*), x \in \cal{X}\}$.  At level $m$, the specified quantiles are renumbered to form $\mathcal{T}_m$ and preserve the ordering of the $\tau_i^*$.  }

{\hy The second set of notation is preferred when describing the probability space that underlies the DQP.  For this, we need to have each variable (with subscripts) refer to a single random element in the construction of the DQP.  These variables appear in different positions in the pyramid.  We introduce the `$\epsilon$-notation' to capture the position in the pyramid and to handle the potentially non-binary splits in the general quantile pyramid.  Consider a pyramid with $m$ levels.  The sequential construction of the pyramid will place the quantile $Q_{x,\epsilon_1}$ at level $1$, $Q_{x,\epsilon_1 \epsilon_2}$ at level $2$, and so on.  The length of the sequence indicates the level at which the quantile was specfied.  The notation also conveys the structural relationship between quantiles.  The parent quantiles of $Q_{x,\epsilon_1 \ldots \epsilon_m}$ are $Q_{x,\epsilon_1 \ldots \epsilon_{m-1}}$ and $Q_{x, \epsilon_1 \ldots (\epsilon_{m-1} + 1)}$.  If there are $K$ quantiles specified in the interval between the parent quantiles, $\epsilon_m$ takes a value in the set $1, \ldots, K$.  The left and right parent quantiles have $\epsilon_m = 0$ and $\epsilon_m = K+1$, respectively.  The quantiles are increasing in $\epsilon_m$.  This results in two distinct sequences $\epsilon_1 \ldots \epsilon_m$ for a quantile that serves as both the left endpoint and the right endpoint of adjacent subintervals.  As the construction proceeds, we use the left-endpoint sequence for which $\epsilon_m = 0$.  For convenience, a short form $\epsilon \equiv \epsilon_1 \cdots \epsilon_{m-1}$ will be used for the $(m-1)$ length sequence throughout the paper.  Figure \ref{fig:notation_DQP} illustrates both notations for a three-level DQP with a given value of $x$.  }


{\hy The oblique DQP is a special case in which all subintervals are split in two.  In this case, $K =0$ or $1$ for all subintervals.  For a dyadic, binary DQP, $K=1$ for all subintervals and the quantile $Q_{x,\epsilon_1 \cdots \epsilon_m}(\tau)$ at level $m$ is the $\tau = \sum_{i=1}^m \epsilon_i 2^{-i}$ quantile of the distribution at $x$.  }

\subsubsection{Definition}
\label{sec:definition}

DQPs come in two main varieties.  The first is the process of finite dependent quantile pyramids (FDQP) while the second, arrived at from a countable sequence of FDQPs, is termed the limit process of dependent quantile pyramids (LDQP).  

We begin with the FDQP.  The FDQP focuses on a finite number of quantiles. 
As with the QP, the FDQP is defined sequentially.  We focus on the quantiles in a single interval (e.g., the center interval at level $2$ in Figure~\ref{fig:notation_DQP}).  The description applies to all such intervals.

\begin{definition}
We call $Q^M = \{Q_x^M, x \in \mathcal{X} \}$ a process of Finite Dependent Quantile Pyramids (FDQP) with $M$ levels if there exists an $M$-level QP valued stochastic process whose distribution $Q^M$ follows. That is, for each $x \in \mathcal{X}$, $Q_x^M$ is an $M$-level QP on $[0,1]$, and, for each $n$ and distinct $x_1, \ldots, x_n \in \mathcal{X}$, Kolmogorov's permutation and marginalization conditions are satisfied.  
\end{definition}

\begin{definition}
We call $F^M = \{F_x^M, x\in\mathcal{X}\}$ a set of conditional distribution functions induced by a process of FDQP with $M$ levels, if for every $x\in\mathcal{X}$, $F_x^M$ is a distribution function and $F_x^M(Q_x^M(\tau)) = \tau$ for all $\tau \in (0,1)$.
\end{definition}

\noindent
The FDQP with $M$ levels is given by its quantile functions, $\{ Q_x^M(\tau), x \in {\cal X}, \tau \in [0, 1] \}$.  It is defined sequentially, beginning with level $1$ of the pyramid.  
Assume that $K$ quantiles are to be drawn at level $m$, for $m = 1, \ldots, M$, in a subinterval that comes from level $m-1$.  The subinterval is $(Q_{x, \epsilon 0}, Q_{x, \epsilon (K+1)})$ for $x \in \mathcal{X}$ and $\epsilon = \epsilon_1 \cdots \epsilon_{m-1}$.  
{\hy The endpoints $Q_{x, \epsilon 0}$ and $Q_{x, \epsilon (K+1)}$ are quantiles inherited from level $m-1$ of the pyramid. When $m=1$, $Q_{x, \epsilon 0} = 0$ and $Q_{x, \epsilon (K+1)}=1$.}
Let $\{V_{x, \epsilon_1 \cdots \epsilon_{m}}, x \in \mathcal{X}\}$ be a multivariate stochastic process with index set ${\cal X}$.  For each $x \in \mathcal{X}$ and $\epsilon = \epsilon_1 \cdots \epsilon_{m-1}$ a random vector $(V_{x, \epsilon 1}, \ldots, 
V_{x, \epsilon (K+1)})$ follows a distribution with the following conditions: (1) $0 < V_{x, \epsilon k} < 1, \text{ for } k = 1, \ldots, (K + 1)$; and (2) $\sum_{k=1}^{K + 1} V_{x, \epsilon k} = 1$.  That is, the vector lies in the interior of the $K$-dimensional simplex.  The quantiles for the $K$ specified quantile levels in the subinterval are
\begin{align}
\label{eq:dqpconstruction}
Q_{x, \epsilon k} &= Q_{x, \epsilon 0} \left(1 - \sum_{j=1}^k V_{x, \epsilon j} \right) + Q_{x, \epsilon (K+1)} \left(\sum_{j=1}^k V_{x, \epsilon j}\right), \quad k = 1, \ldots, K.
\end{align}
A finite number of quantiles, $\tau_1^* < \cdots < \tau_T^*$, is specified in the $M$-level pyramid. 
Quantiles that are not specified directly in the pyramid are given by linear interpolation. For all $\tau \in (\tau_t^*, \tau_{t+1}^*), t = 0, \ldots, T$, we linearly interpolate $Q_{x}^M(\tau)$, i.e.
\[
Q_{x}^M(\tau) = Q_{x}^M(\tau_t^*) + [Q_{x}^M(\tau_{t+1}^*)-Q_{x}^M(\tau_t^*)](\tau - \tau_t^*)/(\tau_{t+1}^* - \tau_t^*). 
\]
The sequential construction and interpolation together define the conditional quantiles given $x$ for all quantile levels $\tau \in (0,1)$.

{\hy
In practice, when determining the number of quantiles, $T$, of an FDQP, we recognize that there is a tradeoff between the number of quantiles considered and computational efficiency.  Our primary recommendation is to focus on the key quantiles of interest.  A secondary, generic recommendation is to prioritize symmetric quantiles such as the three quartiles or the nine deciles.  }

The construction of the FDQP leads to its existence, which is formally {\hy proven in Lemma 3} in Section \ref{sec:dqpexistence}. At each value $x$, the distribution is defined by a finite collection of real valued random variates.  The use of stochastic processes that are well-defined ensures that the QP construction for a single $x$ extends to $\mathcal{X}$.  We restrict attention to cases defined on a single probability space where the FDQP with $M$ levels is the marginal distribution for each of the FDQPs with more than $M$ levels, with appropriate renumbering of the quantiles in $\cup_{m=1}^M \mathcal{T}_m$, where $\mathcal{T}_m$ is the set of quantile levels specified at level $m$.  

With some sufficient conditions, the limit may exist, in which case we have an infinite pyramid, leading to the LDQP. Our concern is with cases where a set of quantiles that is dense in $[0,1]$ is determined in the limit, and so we have no need for interpolation. The existence of the LDQP is established in Section \ref{sec:dqpexistence}. 
When the limit of a sequence of FDQPs exists as $m \to \infty$, the limit process of dependent quantile pyramids (LDQP) is defined.

\begin{definition}
We call $Q = \{Q_x, x \in \mathcal{X} \}$ a limit process of Dependent Quantile Pyramids (LDQP) if there exists a QP valued stochastic process whose distribution $Q$ follows. That is, for each $x \in \mathcal{X}$, $Q_x$ is a QP on $[0,1]$, and, for each $n$ and distinct $x_1, \ldots, x_n \in \mathcal{X}$, Kolmogorov's permutation and marginalization conditions are satisfied.  
\end{definition}

\begin{definition}
We call $F = \{F_x, x\in\mathcal{X}\}$ a set of conditional distribution functions induced by a process of LDQP, if for every $x\in\mathcal{X}$, $F_x$ is a distribution function and $Q_x(\tau) = \inf\{y: F_x(y) \ge \tau\}$ for all $\tau \in (0,1)$.
\end{definition}

\subsection{Canonical Construction}
\label{sec:canonical}
In this section, we provide a canonical construction of the quantiles in {\hy an arbitrary} subinterval at level $m$ of a process of DQPs.  To do so, we construct $U$-processes and $V$-processes and then make use of equation~(\ref{eq:dqpconstruction})  in Section~\ref{sec:dqp}.  We begin under the assumption that the choice of quantiles has been made and that this choice is consistent, matching the correct ordering of the quantiles.  By repeatedly sampling $U$-processes and $V$-processes, we can proceed to sequential construction of the entire process of pyramids.

\subsubsection{U-processes induced from Gaussian processes}

Suppose that we are interested in obtaining $K$ quantiles at level $m$ for a subinterval $(Q_{x, \epsilon 0}, Q_{x, \epsilon (K+1)})$, $m \in \mathbb{N}$.  For each sequence $\epsilon k$ for $k=1, \ldots, K+1$, we consider a Gaussian process (GP) $\{Z_{x, \epsilon k}, x \in \mathcal{X}\}$ with zero mean, unit variance, and some correlation function $\lambda(x, x')$, so that $Z_{x, \epsilon k} \sim \mathcal{GP}(\*0, \lambda(x, x')).$ The correlation function governs the interdependence of quantiles across the covariate space. {\hy We note that the correlation function can differ for different $\epsilon k$. }

We then construct the $U$-processes, $\{U_{x, \epsilon k}, x \in \mathcal{X} \}, k = 1, \ldots, K+1$ for the subinterval, using the normal cumulative distribution function (cdf) transformation element-wise, i.e. for each $x \in \mathcal{X}$,
$U_{x, \epsilon k} = \Phi(Z_{x, \epsilon k}),$
where $\Phi(\cdot)$ denotes the cdf of the standard normal distribution.

\subsubsection{V-processes induced from U-processes via gamma variates}
\label{sec:Vprocesses1}
We construct the $V$-processes, $\{V_{x, \epsilon k}, x \in \mathcal{X} \}, k = 1, \ldots, K+1$, from the $U$-processes.  Let $G_{x, \epsilon k}(\cdot)$ be the gamma distribution function with shape $\alpha_{x, \epsilon k}>0$ and scale $1$, for $k = 1, \ldots, K+1$. For each $x \in \mathcal{X}$ and combination $\epsilon = \epsilon_1 \cdots \epsilon_{m-1}$, define $Y_{x, \epsilon k} \equiv G^{-1}_{x, \epsilon k}(U_{x, \epsilon k}).$ Define $Y_x = \sum_{k=1}^{K+1} Y_{x, \epsilon k}$.  Then, for each $x \in \mathcal{X}$ and $\epsilon$, we set
\begin{equation*}
\label{eq:Vprocess}
(V_{x, \epsilon 1}, \ldots, V_{x, \epsilon (K+1)}) \equiv \left(\frac{Y_{x, \epsilon 1}}{Y_x}, \ldots, \frac{Y_{x, \epsilon (K+1)}}{Y_x}\right)
\end{equation*}
which forms a Dirichlet$(\alpha_{x, \epsilon 1}, \ldots, \alpha_{x, \epsilon (K+1)})$ random vector.  Lastly, collecting $V_{x, \epsilon k}$ together across $x \in \mathcal{X}$, we have constructed a vector-valued process $\{ (V_{x, \epsilon 1}, \ldots, V_{x, \epsilon (K+1)}), x \in \mathcal{X} \}$ with component processes $\{V_{x, \epsilon k}, x \in \mathcal{X}\}$, $k = 1, \ldots, K+1.$  
{\hy When $\alpha_{x, \epsilon k}$ and $G_{x, \epsilon k}$ do not depend on $x$, we omit the subscript $x$ to simpify notation.}  

\subsubsection{V-processes induced from U-processes via beta variates}
\label{sec:Vprocesses2}
The Dirichlet distribution can be derived from the product of independent beta random variates.  This fact leads to a second construction of the $V$-processes.  Define $\alpha_{x, \epsilon 1}, \ldots, \alpha_{x, \epsilon (K+1)}$ as before.  Let $G_{x, \epsilon k}(\cdot)$ be the distribution function of a beta variate with parameters $\alpha_{x, \epsilon k}$ and $\sum_{j=k+1}^{K+1} \alpha_{x, \epsilon j}$.  Define $Y_{x, \epsilon k} = G_{x, \epsilon k}^{-1}(U_{x, \epsilon k})$ for each $x$ and $k = 1, \ldots, K$.  Set 
\begin{eqnarray*}
\label{eq:canonical_beta}
V_{x, \epsilon k} & = & Y_{x, \epsilon k} \prod_{j=1}^{k-1} (1 - Y_{x, \epsilon j}) ,
\end{eqnarray*}
with the conventions that an empty product is $1$ and that $V_{x, \epsilon (K+1)} = 1 - \sum_{k=1}^K V_{x, \epsilon k}$.  The vector $(V_{x, \epsilon 1}, \ldots, V_{x, \epsilon (K+1)})$ follows a Dirichlet$(\alpha_{x, \epsilon 1}, \ldots, \alpha_{x, \epsilon (K+1)})$ distribution.

{\hy
Once $V$-processes are generated, we can proceed to the quantiles $Q_{x, \epsilon 1}, \ldots, Q_{x, \epsilon K}$ in the subinterval, using equation~(\ref{eq:dqpconstruction}).
Given the left and right endpoints of the subinterval, the conditional quantiles at the same quantile level can be expressed as a simple linear transformation of the $V$-processes. This observation also implies the preservation of the dependence structure.
In other words,
\begin{align*}
\text{Corr}(V_{x, \epsilon k}, V_{x', \epsilon k}) = \text{Corr} \left( [Q_{x, \epsilon k} \mid Q_{x, \epsilon 0}, Q_{x, \epsilon (K+1)}], [Q_{x', \epsilon k} \mid Q_{x', \epsilon 0}, Q_{x', \epsilon (K+1)}] \right)
\end{align*}
for any $x, x' \in \mathcal{X}$ and $\epsilon$, and for all $k = 1, \ldots, K$.
When $K=1$ and the construction in this section is used, the random variable $V_{x, \epsilon 1}$ follows a beta distribution.  It is a continuous, monotonic transformation of $U_{x,\epsilon 1}$.  For measures of dependence that are invariant under such a transformation, the (conditional) dependence of $V_{x,\epsilon 1}$ and $V_{x',\epsilon 1}$ matches that of $U_{x,\epsilon 1}$ and $U_{x', \epsilon 1}$.  
}

\subsubsection{Martingale construction}
\label{sec:martingale}

One construction of the dyadic QP in \cite{hjort2009quantile} relies on a martingale.  At each level of the QP, each interval is split in half.  In this canonical construction, $K=1$ and the martingale property requires that $\alpha_{x, \epsilon 1} = \alpha_{x, \epsilon 2}$ for a beta distribution.  Allowing for splits that do not bisect the interval, this becomes, at level $m$ of the pyramid, $\alpha_{x, \epsilon 1} = c_{x, m} (\tau_{\epsilon 1} - \tau_{\epsilon 0})$ and $\alpha_{x, \epsilon 2} = c_{x, m} (\tau_{\epsilon 2} - \tau_{\epsilon 1})$ for some $c_{x, m} > 0$.  

For a non-binary split of an interval, the values of $\alpha_{x, \epsilon k}$ need to be mapped to the relevant subintervals.
{\hy Assume that a subinterval at the $m^{th}$ level contains quantiles associated with $K$ specified quantile levels $\tau_{\epsilon 1} < \ldots < \tau_{\epsilon K}$. Let $\gamma_{\epsilon k}$ for $k=1, \ldots, K$ denote the scaled quantile levels in the subinterval.}
The quantile levels of left and right endpoints of the subinterval are denoted by $\tau_{\epsilon 0}$ and $\tau_{\epsilon (K+1)}$, respectively. Then 
\begin{equation*}
\gamma_{\epsilon k} =  (\tau_{\epsilon k} - \tau_{\epsilon 0})/(\tau_{\epsilon (K+1)} - \tau_{\epsilon 0}), \qquad k=1, \ldots, K.
\end{equation*}
From this, we can derive the values of $\alpha_{x, \epsilon k}$ satisfying the martingale condition. That is, $\alpha_{x, \epsilon k} = c_{x, m}(\tau_{\epsilon k} - \tau_{\epsilon (k-1)})$ for $k=1,\ldots, K+1$ and some $c_{x, m} > 0$.
We note that the chosen quantile levels $\tau_{\epsilon k}$ do not depend on $x$.

\subsection{Quantile Mapping for Transitioning to the Response Space}
\label{sec:mapping}

Assume that the DQP has been constructed on $(0,1)$ and denote the QP at $x$ by $Q_x(\tau)$. As briefly mentioned in \cite{hjort2009quantile}, we can transform the scale as for a proper response space, say, a real line using the inverse normal cdf transformation.  Defining trend parameters $\mu_x$ and scale parameters $\sigma_x$, we have the canonical DQP regression model, centered on a normal theory regression:
\begin{equation}
\label{eq:DQPtransform}
  \{ Q_x^{\mathbb{R}}(\tau) = 
    \underbrace{\mu_x}_\text{Trend} +
    \overbrace{\underbrace{\sigma_x}_\text{Scale} \cdot \Phi^{-1}(Q_x(\tau))}^{\text{Local Fluctuation}}, x \in \mathcal{X} \}.
\end{equation}
The trend parameter $\mu_x$ is shared across the quantiles and controls the overall trend of quantiles throughout the covariate space. The scale parameter controls the dispersion of the distributions while the realized QPs determine the departures from normality.  Together, the scale and QP determine the departure (or local fluctuation) from a set of constant variance normal models with centers given by $\mu_x$.   

Various choices can be made for the trend and scale.  If one believes that there is no significant global trend, $\mu_x$ can be set to a constant.  Alternatively, a linear regression model, $\mu_x = x^\top\*\beta$, may provide an effective choice.  The model is compatible with more complicated models for the trend.  Similarly, models for $\sigma_x$ can range from a simple constant scale to more complex forms.  The trend and scale parameters can be incorporated into Markov chain Monte Carlo (MCMC) procedures used to fit the model.  Alternatively, to save computational effort, estimates can be plugged in and the parameters treated as fixed.

\section{Theoretical Results}
\label{sec:theory}

The novel notation used in this section is formally defined in \nameref{sec:appendix_notation} while proofs of the results appear in \nameref{sec:appendix_proofs}.

\subsection{A Novel Approach to Existence of Quantile Pyramid}
\label{sec:qpexistence}

\cite{hjort2009quantile} provide two different proofs for the existence of the QP. The following results establish the existence of the QP under slightly weaker conditions than those of \cite{hjort2009quantile}.  They also apply to the oblique pyramid of \cite{rodrigues2019pyramid}.

The QP is a probability measure for a distribution-valued random element.  To show its existence, we wish to show that the sequence of probability measures that define the finite quantile pyramids (FQPs) converges to a limiting probability measure.  Our argument relies on two key facts.  First, the space of probability measures over distribution functions with support contained in $[0, 1]$, when equipped with the Prokhorov metric, is compact \citep{parthasarathy1967probability}.  Second, and yet to be established, the sequence of probability measures forms a Cauchy sequence.  Together, these facts lead to the conclusion that the sequence of FQPs converges to a limit and that the limit is a probability measure on distribution-valued elements.  

A first lemma bounds the L\'evy distance, $d_L(\cdot,\cdot)$, between two distribution functions that share a set of quantiles.  Let $0 = \tau_0^* < \tau_1^* < \tau_2^* < \ldots < \tau_T^* < \tau_{T+1}^* = 1$ be ordered quantile levels.  The largest gap between consecutive quantile levels appears in the bound below. 

\begin{lemma}
\label{Lemma:Levy}
Define $\epsilon = \max_{t = 1, \ldots, T+1} (\tau_t^* - \tau_{t-1}^*)$. 
Assume that $F$ and $G$ are two distribution functions such that there exist $y_1, \ldots, y_T$ for which $F(y_t) = G(y_t) = \tau_t^* $ for $ t = 1, \ldots, T$.  Then, $d_L(F,G) \leq \epsilon$.  
\end{lemma}
 
The construction of the QP proceeds through a sequence of FQPs, indexed by $m$, the number of levels in the pyramid.  These FQPs are defined on a single probability space $(\Omega, {\cal B}, \mu)$.  Each $\omega \in \Omega$ defines a sequence of FQPs with $m = 1, 2, \ldots$ levels.  The distributions in the sequence all have support contained in $[0,1]$ and share values of the quantile function at certain specified quantile levels, as described in Section~\ref{sec:generalQP}.  In particular, the $\tau_{\epsilon_1\cdots\epsilon_m}$ quantiles in $F^m(\omega)$ and $F^{m+k}(\omega)$ are identical for all $\omega$ and for any positive integer $k$.  $\mathcal{B}$ is the Borel $\sigma$-field generated by the open sets under the L\'evy metric.

The next lemma shows that the sequence of probability measures arising from the FQPs is Cauchy.   
The measure $\mu_m$ provides the probability distribution on $F^m$.  The distance between $\mu_m$ and $\mu_n$ is measured by the Prokhorov metric, $d_P(\mu_m, \mu_n)$.  

\begin{lemma}
\label{Lemma:Cauchy}
Suppose that $\cup_{m=1}^{\infty} \mathcal{T}_m$ is dense in $[0, 1]$.  Then, under the Prokhorov metric $d_P$, the sequence $\{ \mu_m \}_{m=1}^{\infty}$ is a Cauchy sequence.  
\end{lemma}

A Cauchy sequence on a compact set converges to a limit in the set.  In this case, the sequence of measures $\mu_m$ converges to a limit measure $\mu$.  The limit measure is a probability measure on distribution-valued elements where the distributions have support contained in the interval $[0,1]$.  
This reasoning leads to the existence of the QP, stated formally in the next theorem.  We note that the dyadic construction of the QP satisfies the denseness condition in Lemma~\ref{Lemma:Cauchy}.  

\begin{theorem}
\label{Theorem:existence}
Suppose that $\cup_{m=1}^{\infty} \mathcal{T}_m$ is dense in $[0, 1]$.  Then, the QP constructed as in Section~\ref{sec:quantilepyramid} or \ref{sec:generalQP} is a random element whose distribution is determined by $(\Omega, {\cal B}, \mu)$.    
\end{theorem}

\subsection{Existence of a Process of Dependent Quantile Pyramids}
\label{sec:dqpexistence}

The existence of a process of DQPs follows from consideration of the joint distributions of the QPs at finite sets of indices in the index set.  The construction of the FDQP in Section~\ref{sec:dqp} ensures the existence of the joint distribution of the corresponding sequence of FQPs.  From here, the argument parallels that of the previous section, with the L\'evy metric and Prokhorov metric replaced with their suprema over the finite set of indices.  This ensures the existence of a probability space on which the limiting DQPs satisfy the requisite permutation and marginalization conditions.

The DQP relies on a countable collection of real-valued stochastic processes, $V_{x,\epsilon k}$, all of which have index set ${\mathcal X}$.  Each of these processes satisfies Komogorov's consistency axioms.   The processes are defined on a single probability space.  Selecting a finite set of distinct indices, say, $x_1, \ldots, x_n \in {\cal X}$, the sequence of FDQPs at these indices is generated by a countable collection of real-valued random variables.  The permutation and marginalization conditions follow immediately. 
This establishes the following lemma.

\begin{lemma} (Existence of FDQP)
\label{Lemma:existenceFDQP}
Under the conditions in the previous paragraph, for each $m \in \mathbb{N}$, there exists a distribution-valued stochastic process, $F^m = \{F_x^m, x \in \mathcal{X}\}$, with the specified finite dimensional distributions. 
\end{lemma}

Define the suprema of the L\'evy metric and Prokhorov metric over a set $S$ to be $d_{L_u}(F^m,F^n) = \sup_{x \in S} d_L(F_x^m, F_x^n)$ and $d_{P_u}(\mu^m,\mu^n) = \sup_{x \in S} d_P(\mu_x^m,\mu_x^n)$, respectively.  
Lemma~\ref{Lemma:metric} appearing in \nameref{sec:appendix_proofs} shows that $d_{L_u}$ and $d_{P_u}$ are metrics. 
We first reprise Lemma~\ref{Lemma:Levy}.  

\begin{lemma}
\label{Lemma:Levy2}
Define $\epsilon = \max_{t = 1, \ldots, T+1} (\tau_t^* - \tau_{t-1}^*)$. 
Assume that $F_x$ and $G_x$, $x \in S$, are two sets of distribution functions such that, for each $x \in S$, there exist $y_{x,1}, \ldots, y_{x,T}$ for which $F_x(y_{x,t}) = G_x(y_{x,t}) = \tau_t^*$ for $t = 1, \ldots, T$.  Then, $d_{L_u}(F,G) \leq \epsilon$.  
\end{lemma}

The construction of the FDQPs ensures that, for all $\omega$, for each $x \in \{ x_1, \ldots, x_n \}$ the $\tau_{\epsilon_1\cdots\epsilon_m}$ quantiles in $F_x^m(\omega)$ and $F_x^{m+k}(\omega)$ are identical for every positive integer $k$.  This lets us apply the lemma.  With a suitable choice of quantile levels, the sequence of probability measures for the set of $n$ FQPs is Cauchy.  

\begin{lemma}
\label{Lemma:Cauchy2}
Suppose that $\cup_{m=1}^{\infty} \mathcal{T}_m$ is dense in $[0, 1]$.  Then, with $S = \{ x_1, \ldots, x_n \}$, under the supremum Prokhorov metric $d_{P_u}$, the sequence $\{ \mu_m \}_{m=1}^{\infty}$ is a Cauchy sequence.  
\end{lemma}

Finally, noting that the set of distributions with support contained in $[0,1]^n$ is compact under $d_{L_u}$, by Tychonoff's Theorem, we know that $\mu_m$ converges to some probability measure $\mu$. 
This ensures the existence of the limiting DQP. 

\begin{theorem}
\label{Theorem:existence2}
Suppose that $\cup_{m=1}^{\infty} \mathcal{T}_m$ is dense in $[0, 1]$.  Then, for any $S = \{x_1, \ldots, x_n\} \in \mathcal{X}$,
the DQP for $x \in S$ constructed as in Section \ref{sec:dqp} is a random element whose distribution is determined by $(\Omega, {\cal B}, \mu)$. Furthermore, there exists a stochastic process defined for all $x \in {\cal X}$ whose finite dimensional joint distributions on the QPs are exactly these.   
\end{theorem}

\section{Posterior Inference}
\label{sec:posteriorinference}

In this section, we discuss how to specify the DQP prior  and the likelihood under the canonical construction.  Suppose we have $T$ quantiles of interest, $\tau_1^* < \cdots < \tau_T^*$, to be specified on the DQP.  For simplicity, we work with the binary pyramid.  A similar formulation can be laid out in the general case. Recall that with the canonical construction of the DQP in Section \ref{sec:canonical} and the canonical transformation in Section \ref{sec:mapping}, quantiles are generated in the uniform scale and are transformed to the real line.  
{\hy For notational simplicity, we omit the superscript $\mathbb{R}$ from the quantile in (\ref{eq:DQPtransform}) and let $Q_x(\tau)$ denote the quantiles in the real line scale in this section.}

\subsection{Prior specification}
\label{sec:prior}

A DQP $\{Q_{x}(\tau), \tau \in (0, 1)\}$ is defined for all $x \in \mathcal{X}$ as a stochastic process. Then, the finite-dimensional distributions of the DQP $(\{Q_{x_1}(\tau), \tau \in (0,1)\}, \ldots, \{Q_{x_n}(\tau), \tau \in (0,1)\})$ can be defined for each $n$-tuple $(x_1, \ldots, x_n)$ of distinct elements of $\mathcal{X}$. \citep{billingsley1968}
While the $n$-tuple can be arbitrary, in practice, we choose the coordinates to be covariate values for the data points.

Let the quantiles of interest be denoted by $\*Q$. For convenience, we view $\*Q$ as a $T \times n$ matrix. The $t^{th}$ row of $\*Q$ (denoted by $Q_{\tau_t^*}$) is the vector of ${\tau_t^*}$ quantiles at $x_1, \ldots, x_n$ for $t = 1, \ldots, T$.  The $\tau_t^*$ quantile at $x_i$ is denoted by $Q_{x_i, \tau_t^*}$ for $i=1,\ldots, n$. The parent quantiles that defines the left and right endpoints of the interval on which $Q_{x_i, \tau_t^*}$ is generated are denoted by $Q_{x_i, \tau_t^*}^L$ and $Q_{x_i, \tau_t^*}^R$.  The vectors $Q_{\tau_t^*}^L$ and $Q_{\tau_t^*}^R$ have $Q_{x_i, \tau_t^*}^L$ and $Q_{x_i, \tau_t^*}^R$ as their elements, respectively.

Write $\*\mu_x$ for the vector of trend transformation parameters and $\*\sigma_x$ for the vector of scale transformation parameters, where $\mu_{x_i}$ and $\sigma_{x_i}$ denote the element of $\*\mu_x$ and $\*\sigma_x$, respectively, corresponding to the value of $x_i$.  

{\hy For a GP with zero mean given the input points $(x_1, \ldots, x_n)$, denoted by $\*Z$,  let $\Lambda = [\lambda_{x_i, x_j}]_{i,j=1}^n$ be the correlation matrix, where $\lambda_{x_i, x_j} = \lambda(x_i, x_j)$ is the correlation between $x_i$ and $x_j$ and $\lambda_{x_i, x_i} = 1$.
Let $\Psi(\cdot; a, b)$ denote the cdf of the $beta(a, b)$ distribution and $\Phi(\cdot)$ and $\Phi(\cdot; \eta, \nu)$ denote the cdfs of the $N(0, 1)$ and the $N(\eta, \nu^2)$ distribution, respectively. } 

Given the parameters $\Lambda, \*\mu_x, \*\sigma_x$, the joint conditional density of the finite dimensional distribution of the specified quantile levels of the DQP at $(x_1, \ldots, x_n)$ is
\begin{equation*}
\begin{aligned}
\pi \left(\mathbf{Q} \mid \Lambda, \*\mu_x, \*\sigma_x \right) 
=\prod_{t=1}^{T} \left\{ 
\phi_n\left( h_1(Q_{x_1, \tau_t^*}), \ldots, h_n(Q_{x_n, \tau_t^*}); \*0, \Lambda \right) \times \left\vert \*J(\tau_t^*) \right\vert \right\},
\end{aligned}
\end{equation*}
where $\phi_n(\cdot; \eta, V)$ denotes the probability distribution function (pdf) of the $n$-variate Normal, $N_n(\eta, V)$, the back-transformed quantiles to the $Z$-scale
\begin{align*}
h_i(Q_{x_i, \tau}) &= h_i(Q_{x_i, \tau} | Q_{x_i, \tau}^L, Q_{x_i, \tau}^R, \mu_{x_i}, \sigma_{x_i}) \\
 &= \Phi^{-1}\left(
\Psi \left(\frac{\Phi\left(Q_{x_i, \tau} ; \mu_{x_i}, \sigma_{x_i}\right)-\Phi\left(Q_{x_i, \tau}^L ; \mu_{x_i}, \sigma_{x_i}\right)}{\Phi\left(Q_{x_i, \tau}^R ; \mu_{x_i}, \sigma_{x_i}\right)-\Phi\left(Q_{x_i, \tau}^L ; \mu_{x_i}, \sigma_{x_i}\right)}; a_{\tau}, b_{\tau} \right) 
\right),
\end{align*}
and the determinant of the Jacobian matrix is of the form 
\begin{align*}
|\*J(\tau) | = \prod_{i=1}^n & \left\vert \frac{\partial h_i(Q_{x_i, \tau}| Q_{x_i, \tau}^L, Q_{x_i, \tau}^R)}{\partial Q_{x_i, \tau}}\right\vert\\
= \prod_{i=1}^n & \left\{ \frac{1}{\phi(Z_{x_i, \tau})} \times 
 \psi \left(\frac{\Phi\left(Q_{x_i, \tau} ; \mu_{x_i}, \sigma_{x_i}\right)-\Phi\left(Q_{x_i, \tau}^L ; \mu_{x_i}, \sigma_{x_i}\right)}{\Phi\left(Q_{x_i, \tau}^R ; \mu_{x_i}, \sigma_{x_i}\right)-\Phi\left(Q_{x_i, \tau}^L ; \mu_{x_i}, \sigma_{x_i}\right)}; a_{\tau}, b_{\tau} \right) 
 \right.\\
& \left. \times \frac{\phi\left(Q_{x_i, \tau} ; \mu_{x_i}, \sigma_{x_i}\right)}{\Phi\left(Q_{x_i, \tau}^R ; \mu_{x_i}, \sigma_{x_i}\right)-\Phi\left(Q_{x_i, \tau}^L ; \mu_{x_i}, \sigma_{x_i}\right)} \right\},
\end{align*}
where $\phi(\cdot)$ and $\psi(\cdot; a, b)$ denote the pdfs of the $N(0, 1)$ and the $beta(a, b)$ distribution, respectively.
{\hy The beta distribution parameters $a_{\tau}$ and $b_{\tau}$ can be established using the canonical choice outlined in Section~\ref{sec:canonical}. Given that we are working with the binary pyramid in this section where one quantile is estimated per subinterval, we can denote $\tau_t^*$ as $\tau_{\epsilon 1}$ using the $\epsilon$-notation with an appropriate $\epsilon$ value. For $\tau = \tau_{x, \epsilon 1}$, we set $a_{\tau} = \alpha_{x, \epsilon 1}$ and $b_{\tau} = 1 - \alpha_{x, \epsilon 1}$. }

\subsection{Updating}

Since we use the normal distribution for the transformation from $[0,1]$ to $\mathbb{R}$, the conditional density of the response variable is piecewise-normal. {\hy Given the values of the covariate $\*X = [x_1 \cdots x_n]^\top$ and the parameters $Q$, $\*\mu_x$, and $\*\sigma_x$, the pdf of $\*y$ is }
\begin{align*}
f(\*y | \*Q, \*\mu_x, \*\sigma_x) 
&= \prod_{i=1}^n \left[ \sum_{t=1}^{T+1} \frac{(\tau_t^* - \tau_{t-1}^*)\times I_{(Q_{x_i} (\tau_{t-1}^*), Q_{x_i}(\tau_t^*)]}(y_i) \times \phi(y_i; \mu_{x_i}, \sigma_{x_i}) }{\Phi(Q_{x_i}(\tau_t^*); \mu_{x_i}, \sigma_{x_i}) - \Phi(Q_{x_i}(\tau_{t-1}^*); \mu_{x_i}, \sigma_{x_i})} \right],
\end{align*}
where $I_A(x)$ is an indicator function whose value is one when $x \in A$ and zero otherwise.

\label{sec:mcmc}
{\hy
Utilizing the information present in the data through the likelihood, we can update the prior in Section~\ref{sec:prior} via a MCMC procedure with a Metropolis-Hastings step.  More details can be found in \nameref{sec:mcmc_algorithms}.  If desired, variations on the basic algorithm can be used to improve the mixing of the Markov chain.  
}

\subsection{Inference for the DQP}

{\hy Posterior inference can be made based on the DQP prior and likelihood using the MCMC procedure in \nameref{sec:mcmc_algorithms}.  An estimate of a set of QR curves under squared error loss is obtained by taking the posterior mean of the sample of quantiles at each $x$. }

{\hy The dependence across the covariate space is fundamental to the DQP framework. The conditional quantiles are intrinsically dependent on one another through both the GP covariance structure, $\*\Sigma$, and the pyramid structure.  Typically, increasing the level of dependence results in smoother QR curves, while reducing dependence leads to more flexible and bumpier QR curves. }

\subsubsection{Linearized Inference under the DQP}

Bayesian methods allow one to distinguish between beliefs about a quantity, say a QR curve, and inference about the quantity.  One strategy that has proven successful is to perform modeling in a large space and to impose parsimony through a restriction on the inference (e.g., \cite{maceachern2001decision}, \cite{hahn2015decoupling}).  In contrast, practitioners of classical statistics generally impose parsimony by working in a smaller model space or through model selection.  

Much of the literature on QR, both classical and Bayesian, focuses on linear QR.  The DQP model naturally produces a posterior distribution that assigns probability one to nonlinear QR curves.  To linearize inference, one needs a distribution over the covariate and a set of draws from the posterior distribution of the QR curve.  

{\hy As a criterion, we minimize the integrated squared difference between the nonlinear QR curves and a linear QR curve.  That is, with $x$ following a distribution $G$ and a quantile level of interest $\tau$, and presuming the following integrals are finite, }
\begin{equation}
\label{eq:minimization}
\begin{aligned}
\*\beta_L(\tau) &= \arg\min_{\*\beta^*} \left(\int \int ( Q_{x,\tau} - x^\top\*\beta^*)^2 dP(Q_{x,\tau})dG(x) \right)\\
&= \arg\min_{\*\beta^*} \left(\int ( \expect_x [Q_{x,\tau}] - x^\top\*\beta^*)^2 dG(x) \right),
\end{aligned}
\end{equation}
{\hy where $P(\cdot)$ denotes a posterior distribution function of $Q_{x,\tau}$ and $\expect_x [\cdot]$ denotes the expectation of the quantity at each $x$. }

{\hy In practice, if lacking a distribution $G$, we use the empirical cdf of $x$ for $G$ and the posterior mean of the conditional quantiles to estimate $\expect_x [Q_{x,\tau}]$. With this particular choice, this minimization is obtained as a least squares fit on the posterior mean of the conditional quantiles.  The same $\beta_L(\tau)$ can be obtained in a second fashion.  First, map the nonlinear QR curve at each MCMC iterate to the best fitting linear QR curve with a least squares fit.  Then map this collection of linear QR curves values to the overall best linear QR curve via a second least squares regression of the least squares fits on $x$. More details are contained in \nameref{sec:mcmc_algorithms}.}

\subsubsection{Quantile Prediction for a New $x$ Value}

{\hy 
We can obtain a posterior predictive distribution of quantiles for a new $x$ value, denoted as $x^*$. This is achieved by following a two-step algorithm.  First, we run an MCMC algorithm to sample mapping parameters and DQPs from their joint posterior distribution at the covariate locations where data points exist, namely $x_1, \ldots, x_n$. Next, for a given MCMC iterate, we sample the DQP at the new value $x^*$ from its conditional posterior distribution.

We empasize one important point.  In the event that there is no existing data point at $x^*$, the conditional posterior distribution is just the conditional prior distribution. By collecting the sampled quantiles at the new value $x^*$, we are able to construct the marginal posterior predictive distribution of quantiles for that specific location.
For details on how to perform the posterior sampling of quantiles and mapping parameters from their respective posterior distributions, please refer to Section~\ref{sec:mcmc}.
}

\section{Simulation Study}
\label{sec:simulation} 

We conducted simulation studies to evaluate how the DQP performs.  We considered two sets of quantiles:  $T = 3$ $(0.25, 0.50, 0.75)$ and $T = 7$ $(0.05, 0.10, 0.25, 0.50, 0.75, 0.90, 0.95)$. 
The covariate is $x_{ij} = i$ and the response is $Y_{ij}$ for $i=1, \ldots, 10$ and $j = 1, \ldots, r$.
Let $N(0, 1)$ denote the standard normal distribution and
$t_{df}$ a t-distribution with $df$ degrees of freedom.
We considered the following three scenarios for the simulation study.

\begin{description}
\item[Scenario 1.] (Homogeneous Error) $Y_{ij} = x_{ij} + \epsilon_{ij}, \text{ where } \epsilon_{ij} \overset{iid}{\sim} F $
\item[Scenario 2.] (Heterogeneous Error) $Y_{ij} = x_{ij} + \epsilon_{ij}, \text{ where }  \epsilon_{ij} \overset{iid}{\sim}
\begin{cases}
\sqrt{10} F \hspace{0.5em} \text{ if } 5 \le i \le 6 \\
F \hspace{2.3em} \text{ otherwise }
 \end{cases}$
\item[Scenario 3.] (Nonlinear association between the covariate and the response)\\
(1) $Y_{ij} = \sin(x_{ij}) + \epsilon_{ij}, \text{ where } \epsilon_{ij} \overset{iid}{\sim} N(0, 1)$ \\
(2) $Y_{ij} = \exp(1/x_{ij}) + \epsilon_{ij}, \text{ where } \epsilon_{ij} \overset{iid}{\sim} N(0, 1)$
\end{description}

For Scenarios 1 and 2, (1) $F = N(0,1)$; (2) $F = t_{20}$; (3) $F = t_3$. Scenario 1 examines the case where the error variance is homogeneous and the normality assumption used in the mapping from $(0,1)$ to $\mathbb{R}$ is satisfied (1-1), slightly violated (1-2), and more severely violated (1-3).  In Scenario 2, the error variance is not constant, with a larger variance for certain covariate values, and the mapping is correctly specified (2-1), slightly misspecified (2-2), or more severely misspecified (2-3).  Finally, Scenario 3 showcases settings where the error is normally distributed with homogeneous variance, but the covariate and response variables have a nonlinear association.

For each setting, we used $r = 10$ and $r = 30$ replicates at each value of $x$, corresponding to overall sample sizes of $n = 100$ and $n = 300$, respectively.  In each of our $32$ scenario-$T$-sample size combinations, we generated $N=100$ data sets, fit QRs, and calculated the empirical mean-squared error (MSE) for each quantile $\tau$ at $x \in \mathcal{X}$: $\text{MSE}_x(\tau) = \frac{1}{S} \sum_{s=1}^S (\widehat{Q}_x^s(\tau) - Q_x(\tau))^2,$
where $\hat{Q}_x^s(\tau)$ is an estimated quantile for $\tau$ level with $s^{th}$ simulated data set at $x$.  We further computed the MSE for the quantile, averaged over the distribution of $x$ as $\text{MSE}(\tau) = \sum_{x=1}^{10} 0.1 \,\, \text{MSE}_x(\tau)$, which again is averaged over $\tau$ as $\text{AMSE} = \sum_{t=1}^T \sum_{x=1}^{10} 0.1 \,\, \text{MSE}_x(\tau)/T$.

We fit a binary FDQP with two levels for the $T=3$ case and with three levels for the $T=7$ case based on the normal inverse cdf transformation from (\ref{eq:DQPtransform}), centering the FDQP on a linear regression model with $\mu_x = x^\top \*\beta$.
{\hy Regarding the dependence of the pyramids between locations $x$ and $x'$ in the covariate space, we used the canonical construction for the Gaussian processes with zero mean vector and Gaussian correlation function $\lambda(x, x') = \exp(-\Vert x - x' \Vert^2 / \phi) $ with $\phi=5$.
For the beta distribution at level $m$ in the construction of the binary pyramid prior, we used $\alpha_{\epsilon 1} = c_m (\tau_{\epsilon 1} - \tau_{\epsilon 0})$ and $\alpha_{\epsilon 2} = c_m (\tau_{\epsilon 2} - \tau_{\epsilon 1})$ with $c_m = (m+5)^2$.
The prior distribution for $\boldsymbol{\beta}$ is bivariate normal with mean $\mu_0 = (5, 0)^\top$ and variance matrix $\Sigma_0 = \mbox{diag}(3, 3)$. }
For the scale transformation parameter $\sigma_x$, we used the sample standard deviation at $x$ as a plug-in estimator. 

We compare our method to four alternatives: {\hy a classical QR approach by \cite{koenker1978regression} (`quantreg'), Yu \&  Moyeed's Bayesian QR approach implemented by \cite{benoit2017bayesqr} (`bayesQR'), a simultaneous linear QR by \cite{yang2017joint} (`qrjoint'), and its generalized variant incorporating spatial dependence using a Gaussian process by \cite{chen2021joint} (`JSQR-GP')}. The estimators are evaluated by \text{AMSE}.  For the MCMC runs of bayesQR, qrjoint, JSQR-GP, and DQP, we used a warmup of $1,000$ iterates followed by $100,000$ draws for estimation, thinned at a rate of $1$ in $100$.

The AMSE values are summarized in Figure~\ref{fig:AMSE}, while the corresponding values and standard errors can be found in Table~\ref{table:MSEn100} and Table~\ref{table:MSEn300} in \nameref{sec:appendix_tables}.
Examining Figure~\ref{fig:AMSE},
the linearized DQP (`DQP-lm') demonstrated competitive performance relative to the alternative approaches in Scenario 1. 
{\hy It either had the lowest AMSE or had an AMSE within a two standard error range of the lowest value for all of the cases of Scenarios 1-1 and 1-2 except for the $T=7$ and $n=100$ case.} This suggests that when the underlying relationship between the response and covariates is believed to be linear, the DQP can be effectively used for linear inference, yielding reasonable results.
{\hy In Scenario 1, qrjoint had the lowest AMSE in most cases, primarily because its linearity assumption in the model closely aligns with the underlying truth and it simultaneously estimates multiple conditional quantiles.
We note that DQP struggled when the normal assumption in scale transformation of DQP is severely violated (Scenario 1-3).}

Under Scenarios 2-1 and 2-2, the DQP ourperformed all other methods. This is because the DQP is not only robust to mild misspecification of the mapping from $(0, 1)$ to the response variable scale but can also account for the local nonlinearity arising from a larger variance at specific locations of the covariate. Among the linear methods, DQP-lm performed the best. {\hy This observation may be attributed to the fact that DQP-lm identifies linear QR curves by leveraging the QR curves estimated within the nonlinear model space.}
Even under Scenario 2-3, where the normal assumption in the DQP transformation is violated, the DQP still outperformed other methods in most of the cases except when $n=100$ and $T=7$. 

Regarding the nonlinear association between the covariate and the response, represented by Scenarios 3-1 and 3-2, the DQP has lower AMSE than the other alternatives for most combinations of $n$ and the number of quantiles, except for the case of $n=100$ and $T=7$ under Scenario 3-2. 
This is because the DQP can capture the local nonlinearity of quantiles, such as the cyclic movement in Scenario 3-1 and the sudden drop for small covariate values in Scenario 3-2. 
The advantage of the DQP becomes more evident as $n$ increases.

\section{Application}
\label{sec:cyclone}
\cite{elsner2008increasing} observed that the trend in tropical cyclone intensity in the North Atlantic Ocean from 1981 to 2006 is different for different quantiles. They fit separate linear QRs for multiple quantiles using the R package `quantreg' \citep{koenker_2005} and found that the higher quantiles showed an upward trend, with statistically significant slopes above the 0.7 quantile. Lower quantiles had slopes closer to zero.

\cite{kadane2012simultaneous} analyzed the same data set and came to a different conclusion.
They developed a Bayesian approach to simultaneously fit a collection of linear QRs. 
They found that the increasing trend in the cyclone intensity was significant for almost all quantiles, not just the upper quantiles. 

We analyzed the data set (https://myweb.fsu.edu/jelsner/temp/Data.html) from \cite{elsner2008increasing}. 
The data consist of the lifetime maximum wind speeds of $291$ North Atlantic tropical cyclones derived from satellite imagery.  The wind speeds are the response ($Y$) and range from 29.8 to 159.5 $ms^{-1}$.  The year is the covariate ($x$). 

We fit a binary DQP with four levels and 15 quantiles at $\tau$ = 0.05, 0.10, 0.20, 0.25, 0.30, 0.35, 0.40, 0.50, 0.60, 0.65, 0.70, 0.75, 0.80, 0.90, and 0.95.
{\hy We used the binary canonical construction with Gaussian processes with zero mean vector and the exponential correlation function $\lambda(x, x') = \exp(-\Vert x - x' \Vert / \phi) $ with $\phi=5$.}
{\hy For the beta distribution at level $m$ in the construction of $V$ processes, we used $\alpha_{\epsilon 1} = c_m (\tau_{\epsilon 1} - \tau_{\epsilon 0})$ and $\alpha_{\epsilon 2} = c_m (\tau_{\epsilon 2} - \tau_{\epsilon 1})$ with $c_m = (m+5)^2$.}
We again used the canonical transformation with linear regression model for the trend parameter, i.e. $\mu_x = x^\top \*\beta$.
The prior distribution for $\boldsymbol{\beta}$ is bivariate normal with mean $\mu_0 = (75, 0.5)^\top$ and variance matrix $\Sigma_0 = \mbox{diag}(15, 2)$.  
{\hy For the scale transformation parameter $\sigma_x$, we fitted an ordinary least squares model with the sample standard deviation regressed on each year and then used the resulting fitted values.} 
We ran $10,000$ warm up iterates of MCMC followed by $200,000$ iterates, thinned to $2,000$ iterates for estimation.  
{\hy For the purpose of comparison, we also applied the alternative simultaneous QR methods discussed in Section~\ref{sec:simulation} to the same dataset. }

{\hy Figure~\ref{fig:cyclone_plots2} displays the analysis results from the simultaneous QR methods, including our approach.}
In panel (a), the grey lines provide the posterior means at each year and the black lines the linearized fit. The lines are overlaid on the data points. 
The 95\% credible intervals for the slopes of the linearized fit are shown in the right panel of (a). 
The slopes tend to be greater for greater quantiles, implying that stronger cyclones have been getting stronger more quickly. Indeed, the 95\% credible interval for the slope of the 0.95 quantile is observed to be well above that of the 0.50 quantile. This means that the most intense cyclones are increasing in strength at a significantly faster rate than the moderately strong cyclones. The 95\% credible intervals for the slope of the quantiles below 0.50 include zero. The slopes do not significantly differ from zero for lower quantiles. 
Overall, our method applied to these data shows the intensity of the upper half of the cyclones to be increasing, with more powerful cyclones intensifying at a more rapid rate. 

{\hy 
On the other hand, the outcome obtained from qrjoint in panel (b) reveals that all the 95\% credible intervals lie above zero while overlapping with one another. This suggests that cyclone intensities are increasing across all levels, although the rate of increase may not significantly differ among cyclones with varying power. Meanwhile, the findings from JSQR-GP in panel (c) indicate that only the slopes for the strongest cyclones have values greater than zero, with their respective intervals also overlapping. While we lack knowledge of the absolute ground truth, it becomes evident that all three models concur on one point: the intensity of the strongest cyclones is increasing.
}

\section{Discussion}
\label{sec:discussion}

We propose a nonparametric Bayesian approach to quantile regression, based on a process of DQPs.  The DQP generalizes {\hy the QP of \cite{hjort2009quantile}} to allow dependence across a predictor space.  The flexibility of the model allows us to depart from linearity and to handle regressions for multiple quantiles in unbounded predictor spaces without quantile crossing.  

The canonical construction can be adapted to account for a various features of the data.  As examples, the standard normal distribution used in the inverse cdf transformation can be replaced with a distribution with different tails to handle thick or thin tailed data; skewness in the quantiles can be handled by replacing the symmetric normal distribution with an asymmetric distribution; and positively valued responses can be handled by basing the transformation on a distribution supported on the half line.  The linear regression model for the mean can be replaced with a nonlinear form, resulting in large-scale nonlinearity of the QRs.  Simple choices for this include the use of a deterministic form such as a fixed-knot spline or a stochastic form such as a Gaussian process.  
Replacement of the linear form for the scale factor ($x^\top \*\gamma$) with a form that ensures positivity, for example, $\exp( x^\top \*\gamma)$, relieves concerns about the implications of unbounded $\mathcal{X}$.  

In the simulation examples, we employed the sample standard deviation at the location $x$ as a plug-in estimator for the scale transformation parameter $\sigma_x$. An alternative is to use a pooled sample standard deviation in situations where heterogeneity is not suspected, such as Scenario 1. The results of this alternative will be included in our future work. 
The simulations we report above rely on a single predictor.  The stochastic processes used for the FDQP are Gaussian processes that are then passed through transformations to arrive at the FDQP.  Extension to the case of multiple predictors is straightforward through the use of Gaussian processes with a multivariate index.  The Gaussian processes can be replaced with other processes.

\section*{Acknowledgements}
The authors wish to thank the reviewers for comments that improved the paper.  The authors gratefully acknowledge support from National Science Foundation grants DMS-2015552 and SES-1921523.

\section*{Disclosure statement}
No potential conflict of interest was reported by the author(s).

\bibliographystyle{agsm}
\bibliography{bibfile}

\section{Appendix A}
\label{sec:appendix_notation}

\begin{itemize}
\item $(\Omega, \mathcal{B}, \mu)$ := a triple that defines a probability space
\item $D$ = the space of all distributions with the support on $[0, 1]$
\item $D_S$ = the product space of all collections of (conditional) distributions with the support on $[0, 1]$ on some finite set $S \subset \mathcal{X}$
\item $\mathcal{B}'$ := the Borel $\sigma$-field of $D$, i.e. all Borel sets of distributions on $D$
\item $\mathcal{B}'_S$ := the Borel $\sigma$-field of $D_S$
\item $d_L(F,G)$ := Lévy metric between distributions $F$ and $G$ ($F,G \in D$)
\item $d_L(F,A) = \inf \{d_L(F,G) \mid G \in A \}$ := Lévy metric between a distribution $F \in D$ and a Borel set $A \in \mathcal{B}'$ 
\item $A_\alpha = \{F \in D \mid d_L(F,A) < \alpha \}$ := the open $\alpha$ ball ($\alpha > 0$) about $A \in \mathcal{B}'$ \\ (or $A_\alpha = \{F \in D \mid d_{L_u}(F,A) < \alpha \}$ for $A \in \mathcal{B}'_S$)
\item $\mathcal{P} = \mathcal{P}(\mathcal{B}')$ := all Borel probability measures on the Borel $\sigma$-field of $D$
\item $\mathcal{P}_{S} = \mathcal{P}_{S}(\mathcal{B}'_{S})$ := all Borel probability measures on the Borel $\sigma$-field of $D_{S}$
\item $d_P(\mu,\nu)$ := the Prokhorov metric on $\mathcal{P}$ for $\mu, \nu \in  \mathcal{P}$, that is
\[
d_P(\mu,\nu) = \inf_{\alpha > 0} \{ \alpha \mid \mu(A) \leq \nu(A_\alpha) + \alpha \mbox{ and } \nu(A) \leq \mu(A_\alpha) + \alpha \mbox{ for all } A \in \mathcal{B}' \}
\]

\item $F^m$ := an $m$-level FQP in Section \ref{sec:qpexistence} and FDQP in Section \ref{sec:dqpexistence} 
\item $\Omega_{A,m} = \{ \omega \mid F^m(\omega) \in A \}$ := an $\omega$-set with $F^m$ and $A \in \mathcal{B}'$
\item $\Omega_{A_\alpha,m} = \{ \omega \mid F^m(\omega) \in A_\alpha \}$ := the $\alpha$ ball for $F^m(\omega)$ about $A \in \mathcal{B}'$ 
\item $\mu_m$ := a probability measure in $\mathcal{P}$ arising from all the $m$-level pyramids, which assigns probabilities to subsets of $D$ (or $D_S$ for some $S 
\subset \mathcal{X}$)
\item $\mu_m(A) = \mu(\Omega_{A,m})$ := the probability that the probability distributions induced from $m$-level pyramids belong to $A \in \mathcal{B}'$
\end{itemize}

\section{Appendix B}
\label{sec:appendix_proofs}

\noindent {\bf Proof of Lemma~\ref{Lemma:Levy}.}
Consider $y \in [\tau_{t-1}^*, \tau_t^*]$.  Then $\tau_{t-1}^* \leq F(y), G(y) \leq \tau_t^*$.  Thus $F(y) - \epsilon \leq \tau_t^* - \epsilon \leq \tau_{t-1}^*$ and so $F(y) - \epsilon \leq G(y)$.  Similarly, $F(y) + \epsilon \geq \tau_{t-1}^* + \epsilon \geq \tau_t^*$ and so $F(y) + \epsilon \geq G(y)$.  Thus $F(y - \epsilon) - \epsilon \leq G(y) \leq F(y + \epsilon) + \epsilon$.  
Repeating the argument above for $t = 1, \ldots, T+1$ establishes that $d_L(F,G) \leq \epsilon$.  
$\blacksquare$

\vspace{1em}

\noindent {\bf Proof of Lemma~\ref{Lemma:Cauchy}.}
Fix $\epsilon > 0$.  Choose $M$ such that $\max_{t \in \{1, \ldots, T_M+1\}} (\tau_t^* - \tau_{t-1}^*) < \epsilon$.  Applying Lemma~\ref{Lemma:Levy}, we have $d_L(F^m(\omega), F^n(\omega)) < \epsilon$ for all $m, n \geq M$ and for all $\omega \in \Omega$.  That is, $F^n(\omega)$ is in the $\epsilon$-ball about $F^m(\omega)$.  Such an $M$ always exists provided that $\cup_{m=1}^{\infty}\mathcal{T}_m$ is dense in $[0, 1]$ and each $F^m(\omega)$ has the same pyramid structure.

Consider an arbitrary Borel set $A \in \mathcal{B}'$.  Define the sets $\Omega_{A,m} =  \{ \omega \mid F^{m}(\omega) \in A \}$ and $\Omega_{A_\epsilon,n} = \{ \omega \mid F^{n}(\omega) \in A_\epsilon \}$.  For every $\omega \in \Omega_{A,m}$, $F^n(\omega)$ is in the $\epsilon$-ball about $F^m(\omega)$.  Hence $\omega \in \Omega_{A_\epsilon,n}$ and so $\Omega_{A,m} \subset \Omega_{A_\epsilon,n}$.  

Turning to the probability measures on the FQPs with $m$ and $n$ levels, we note that $\mu_m(A) = \mu(\Omega_{A,m})$ and $\mu_n(A_\epsilon) = \mu(\Omega_{A_\epsilon,n})$.  Since $\Omega_{A,m} \subset \Omega_{A_\epsilon,n}$, $\mu_m(A) \leq \mu_n(A_\epsilon) < \mu_n(A_\epsilon) + \epsilon$. A similar argument shows that $\mu_n(A) < \mu_m(A_\epsilon) + \epsilon$. This holds for all Borel $A$, and so $d_P(\mu_m, \mu_n) \leq \epsilon$.  Thus, for each $\epsilon > 0$, there is an $M$ such that, for all $m, n \geq M$, $d_P(\mu_m, \mu_n) \leq \epsilon$.  The sequence $\{ \mu_m \}_{m=1}^\infty$ is Cauchy.  $\blacksquare$

\vspace{1em}

\noindent {\bf Proof of Theorem~\ref{Theorem:existence}.}
By Lemma~\ref{Lemma:Cauchy}, the sequence $\{\mu_m\}_{m=1}^{\infty}$ is Cauchy. 
Moreover, the space $\mathcal{P}$ equipped with the Prokhorov metric is compact, and thus complete, since the space $D$ equipped with Lévy metric is compact (see Theorem 2.6.4 in \cite{parthasarathy1967probability}). Therefore, the sequence $\{\mu_m\}_{m=1}^{\infty}$ is convergent. That is, there exists $\mu$ to which $\mu_m$ converges and that $\mu$ provides a probability distribution on $\lim_{m\to \infty} F^m$. Thus, a limit of QP exists. $\blacksquare$ 

\vspace{1em}
\noindent

\begin{lemma}
\label{Lemma:metric}
$d_{L_u}$ as defined in Section~\ref{sec:dqpexistence} is a metric on the space of distributions.  $d_{P_u}$ is a metric on the space of probability measures over distributions.  The set $\cal{S}$ need not have finite cardinality.  
\end{lemma}

\noindent{\bf Proof.}
Symmetry, non-negativity, the triangle inequality, and the zero property follow from straightforward calculation.  
With a nod to the St.\ Petersburg paradox, we must show that $d_{L_u}(F^m,F^n) < \infty$.  Since $d_L(F_x^m,F_x^n) \leq 1$ for all $x \in {\cal S}$, we have that $d_{L_u}(F^m,F^n) \le 1$.  The argument for $d_{P_u}$ is established in the same way.  $\blacksquare$

\vspace{1em}
\noindent {\bf Proof of Lemma~\ref{Lemma:Levy2}.}
From Lemma~\ref{Lemma:Levy}, we have $d_L(F_x,G_x) \leq \epsilon$ for all $x \in {\cal S}$.  Thus $d_{L_u}(F,G) \leq \epsilon$.  $\blacksquare$

\vspace{1em}
\noindent {\bf Proof of Lemma~\ref{Lemma:Cauchy2}.}
Replace  $d_L$ with $d_{L_u}$, $d_P$ with $d_{P_u}$, and $\mathcal{B}'$ with $\mathcal{B}'_S$ in the proof of Lemma~\ref{Lemma:Cauchy}.  $\blacksquare$

\vspace{1em}
\noindent{\bf Proof of Theorem~\ref{Theorem:existence2}.}
The proof of the first part of the theorem follows that of Theorem~\ref{Theorem:existence}.  
Kolmogorov's permutation condition is satisfied at each step in the sequence of FDQPs since each of the $V_x$ processes satisfies the condition.  His marginalization condition is also satisfied.  Since $\{x_1, \ldots, x_n \}$ was arbitrary, this ensures the existence of a stochastic process with the specified limiting distributions (e.g., \cite{billingsley1968}, chapter 7).  $\blacksquare$

\clearpage
\section{Appendix C}
\label{sec:mcmc_algorithms}

\subsection{MCMC Algorithms}
{\hy
\begin{enumerate}
\item Sample $\*Q = [ Q_{\tau_1}, \cdots, Q_{\tau_T}]^\top$ given $\*\mu_x$ and $\*\sigma_x$
\begin{enumerate}[(i)]
\item Choose $\tau_t$ following the top-down pyramid structure.
\item Sample $Q_{\tau_t}^p = (Q_{x_1, \tau_t}^p, \cdots, Q_{x_n, \tau_t}^p)^\top$ given the neighboring quantiles $Q_{\tau_{t-1}} = (Q_{x_1, \tau_{t-1}}, \cdots, Q_{x_n, \tau_{t-1}})^\top$ and $Q_{\tau_{t+1}} = (Q_{x_1, \tau_{t+1}}, \cdots, Q_{x_n, \tau_{t+1}})^\top$, for $t = 1, \ldots, T$ from a conditional proposal density $q(\cdot \mid Q_{-\tau_t}, \*\mu_x, \*\sigma_x)$.
\item Accept the new proposal $\*Q^p$ whose $t^{th}$ row is replaced with $Q_{\tau_t}^p$ over the current value $\*Q^c$  with probability 
$$\alpha = \min\left\{1, \frac{\pi(\*Q^p \mid \*\mu_x, \*\sigma_x) f(\*y \mid \*Q^p, \*\mu_x, \*\sigma_x)}{\pi(\*Q^c \mid \*\mu_x, \*\sigma_x) f(\*y \mid \*Q^c, \*\mu_x, \*\sigma_x)} \times \frac{q(Q_{\tau_t}^c \mid Q_{-\tau_t}, \*\mu_x, \*\sigma_x)}{q(Q_{\tau_t}^p \mid Q_{-\tau_t}, \*\mu_x, \*\sigma_x)} \right\}.$$
\item Repeat (i) - (iii) until the bottom of the pyramid.
\end{enumerate}
\item Sample $\*\mu_x$ given $\*Q$ and $\*\sigma_x$
\begin{enumerate}[(i)]
\item Partition $\*\mu_{x_1}, \ldots, \*\mu_{x_n}$ into $I$ blocks.  Call block $i$ $\*\mu_x^{i}$.  
\item Sample $\mu_x^{(i), p}$ from a conditional proposal density $q(\cdot \mid \mu_x^{(i), c})$ and accept with probability 
$$\alpha = \min\left\{1, \frac{\pi(\mu_x^{(i), p} \mid \*\mu_x^{-(i)}) f(\*y \mid \*Q, \*\mu_x^p, \*\sigma_x)}{\pi(\mu_x^{(i), c} \mid \*\mu^{-(i)}) f(\*y \mid \*Q, \*\mu_x^c, \*\sigma_x)} \times \frac{q(\pi(\mu_x^{(i), c} \mid \mu_x^{(i), p})}{q(\pi(\mu_x^{(i), p} \mid \mu_x^{(i), c})} \right\}.$$
\item Repeat step (ii) for $i = 1, \ldots, I$.  
\end{enumerate}
\item Sample $\*\sigma_x$ given $\*Q$ and $\*\mu_x$ following a similar step to 2. 
\end{enumerate}
Note that the sampling of $\*Q$ can be further broken down to sampling of smaller blocks as in step 2.  Doing so impacts the acceptance rate of the proposals and the mixing of the Markov chain.  
}

\subsection{Linearized Inference}
{\hy Throughout this derivation, we assume that the integrals are finite.  This ensures the non-trivial existence of a minimizing $\*\beta^*$.  
\begin{equation*}
\begin{aligned}
\*\beta_L(\tau) &= \arg\min_{\*\beta^*} \left(\int \int ( Q_{x,\tau} - x^\top\*\beta^*)^2 dP(Q_{x,\tau})dG(x) \right)\\
&=\arg\min_{\*\beta^*} \left(\int \int ( Q_{x,\tau} - \expect_x [Q_{x,\tau}] + \expect_x [Q_{x,\tau}] - x^\top\*\beta^*)^2 dP(Q_{x,\tau})dG(x) \right)\\
&=\arg\min_{\*\beta^*} \left(
\int \int( Q_{x,\tau} - \expect_x [Q_{x,\tau}])^2 dP(Q_{x,\tau})dG(x) 
+ \int \int(\expect_x [Q_{x,\tau}] - x^\top\*\beta^*)^2 dP(Q_{x,\tau})dG(x) \right. 
\\ 
& \qquad \qquad + \left. \int \int 2 ( Q_{x,\tau} - \expect_x [Q_{x,\tau}]) (\expect_x [Q_{x,\tau}] - x^\top\*\beta^*)      dP(Q_{x,\tau})dG(x) \right) \\
&= \arg\min_{\*\beta^*} \left( \int ( \expect_x [Q_{x,\tau}] - x^\top\*\beta^*)^2 dG(x) \right),
\end{aligned}
\end{equation*}
where $P(\cdot)$ denotes the posterior distribution function of $Q_{x,\tau}$, $G(\cdot)$ denotes the distribution function of $x$, and $\expect_x [\cdot]$ denotes the expectation of the quantity at each $x$.
Moreover, if $G(x)$ is the empirical cdf of $x$, then the problem above becomes
\begin{equation*}
\begin{aligned}
\*\beta_L(\tau) &= \arg\min_{\*\beta^*} 
\left(\sum_{i=1}^n \frac{1}{n} ( \expect_x [Q_{x,\tau}] - x_i^\top\*\beta^*)^2  \right),
\end{aligned}
\end{equation*}
which is equivalent to the least squares linear regression problem. }

\clearpage
\section{Appendix D}
\label{sec:appendix_tables}
\begin{table}[h!]
\centering
{\renewcommand{\arraystretch}{1.7}
\scriptsize
\begin{tabular}{ x{4.5em} | x{1em} | x{4em} x{4em} x{4em} x{5em} x{4em} x{4.2em} } 
Scenario & T & quantreg & bayesQR & qrjoint & JSQR-GP & DQP & DQP-lm 
\tn
\hline
1-1 & 3 & 0.0353 & 0.0293 & {\bf 0.0277} & 0.0314 & 0.0463 & 0.0301 \tn
& & (0.0035) & (0.0029) & (0.0028) & (0.0038) & (0.0032) & (0.0031) \tn
 & 7 & 0.0574 & 0.0421 & {\bf 0.0350} & 0.0405 & 0.1076 & 0.0510 \tn
& & (0.0055) & (0.0042) & (0.0034) & (0.0042) & (0.0058) & (0.0051) \tn
\hline
1-2 & 3 & 0.0395 & 0.0323 & {\bf 0.0304} & 0.0328 & 0.0540 & 0.0331 \tn
& & (0.0035) & (0.0030) & (0.0028) & (0.0029) & (0.0035) & (0.0033) \tn
 & 7 & 0.0702 & 0.0506 & {\bf 0.0404} & 0.0461 & 0.1293 & 0.0530 \tn
& & (0.0076) & (0.0054) & (0.0039) & (0.0044) & (0.0064) & (0.0050) \tn
\hline
1-3 & 3 & 0.0541 & 0.0449 &{\bf  0.0448} & 0.0534 & 0.2043 & 0.0959 \tn
& & (0.0066) & (0.0055) & (0.0049) & (0.0059) & (0.0176) & (0.0108) \tn
 & 7 & 0.2531 & 0.1551 & {\bf 0.1113} & 0.1788 & 0.6842 & 0.1712 \tn
& & (0.0321) & (0.0180) & (0.0135) & (0.0241) & (0.0642) & (0.0195) \tn
\hline
2-1 & 3 & 0.2937 & 0.2809 & 0.2788 & 0.2809 & {\bf 0.0861} & 0.2636 \tn
& & (0.0059) & (0.0049) & (0.0050) & (0.0085) & (0.0058) & (0.0038) \tn
 & 7 & 1.3685 & 1.1694 & 1.1271 & 1.2326 & {\bf 0.2460} & 1.1033 \tn
& & (0.0511) & (0.0187) & (0.0117) & (0.0321) & (0.0166) & (0.0080) \tn
\hline
2-2 & 3 & 0.3021 & 0.2917 & 0.2888 & 0.2844 & {\bf 0.1041} & 0.2762 \tn
& & (0.0061) & (0.0056) & (0.0054) & (0.0049) & (0.0069) & (0.0041) \tn
 & 7 & 1.4621 & 1.2579 & 1.2129 & 1.3279 & {\bf 0.2958} & 1.1887 \tn
& & (0.0525) & (0.0170) & (0.0110) & (0.0287) & (0.0185) & (0.0073) \tn
\hline
2-3 & 3 & 0.3832 & 0.3679 & {\bf 0.3641} & 0.3828 & 0.4922 & 0.4434 \tn
& & (0.0087) & (0.0073) & (0.0068) & (0.0112) & (0.0944) & (0.0206) \tn
 & 7 & 2.8263 & 2.2268 & 2.1619 & 2.4989 & {\bf 1.7753} & 2.1811 \tn
& & (0.1651) & (0.0434) & (0.0390) & (0.0885) & (0.3882) & (0.0539) \tn
\hline
3-1 & 3 & 0.5219 & 0.5192 & 0.5099 & 0.5486 & {\bf 0.3356} & 0.5105 \tn
& & (0.0053) & (0.0048) & (0.0041) & (0.0074) & (0.0063) & (0.0041) \tn
 & 7 & 0.5869 & 0.6074 & 0.5684 & 0.6910 & {\bf 0.4031} & 0.5253 \tn
& & (0.0116) & (0.0104) & (0.0080) & (0.0150) & (0.0094) & (0.0055) \tn
\hline
3-2 & 3 & 0.1365 & 0.1302 & 0.1292 & 0.1384 & {\bf 0.1134} & 0.1324 \tn
& & (0.0035) & (0.0033) & (0.0030) & (0.0042) & (0.0037) & (0.0030) \tn
 & 7 & 0.1661 & 0.1633 & {\bf 0.1408} & 0.1615 & 0.1756 & 0.1526 \tn
& & (0.0071) & (0.0063) & (0.0043) & (0.0061) & (0.0068) & (0.0049) \tn
\hline
\end{tabular}
}
\caption{AMSE values over 100 simulated datasets with standard errors in parentheses when $n=100$. }
\label{table:MSEn100}
\end{table}

\clearpage
\begin{table}[h!]
\centering
{\renewcommand{\arraystretch}{1.7}
\scriptsize
\begin{tabular}{ x{4.5em} | x{1em} | x{4em} x{4em} x{4em} x{5em} x{4em} x{4.2em} } 
Scenario & T & quantreg & bayesQR & qrjoint & JSQR-GP & DQP & DQP-lm 
\tn
\hline
1-1 & 3 & 0.0116 & 0.0105 & 0.0100 & 0.0113 & 0.0175 & {\bf 0.0092} \tn
& & (0.0012) & (0.0011) & (0.0010) & (0.0012) & (0.0010) & (0.0009) \tn
 & 7 & 0.0183 & 0.0148 & {\bf 0.0133} & 0.0144 & 0.0367 & 0.0151 \tn
& & (0.0019) & (0.0016) & (0.0013) & (0.0015) & (0.0018) & (0.0014) \tn
\hline
1-2 & 3 & 0.0134 & 0.0121 & {\bf 0.0105} & 0.0120 & 0.0195 & 0.0106 \tn
& & (0.0014) & (0.0013) & (0.0010) & (0.0011) & (0.0012) & (0.0010) \tn
 & 7 & 0.0239 & 0.0199 & {\bf 0.0153} & 0.0171 & 0.0417 & 0.0176 \tn
& & (0.0024) & (0.0021) & (0.0015) & (0.0017) & (0.0022) & (0.0016) \tn
\hline
1-3 & 3 & 0.0179 & 0.0163 & {\bf 0.0147} & 0.0269 & 0.0781 & 0.0343 \tn
& & (0.0017) & (0.0017) & (0.0014) & (0.0092) & (0.0075) & (0.0036) \tn
 & 7 & 0.0765 & 0.0587 & {\bf 0.0401} & 0.0964 & 0.3621 & 0.1015 \tn
& & (0.0086) & (0.0065) & (0.0041) & (0.0282) & (0.0504) & (0.0133) \tn
\hline
2-1 & 3 & 0.2593 & 0.2571 & 0.2561 & 0.4969 & {\bf 0.0434} & 0.2409 \tn
& & (0.0026) & (0.0026) & (0.0024) & (0.0922) & (0.0035) & (0.0015) \tn
 & 7 & 1.1491 & 1.1068 & 1.0972 & 1.3953 & {\bf 0.0965} & 1.0521 \tn
& & (0.0162) & (0.0103) & (0.0107) & (0.0750) & (0.0072) & (0.0026) \tn
\hline
2-2 & 3 & 0.2714 & 0.2681 & 0.2654 & 0.4406 & {\bf 0.0454} & 0.2503 \tn
& & (0.0030) & (0.0031) & (0.0025) & (0.0901) & (0.0039) & (0.0014) \tn
 & 7 & 1.2422 & 1.1992 & 1.1776 & 1.4726 & {\bf 0.0986} & 1.1391 \tn
& & (0.0193) & (0.0122) & (0.0071) & (0.0728) & (0.0064) & (0.0025) \tn
\hline
2-3 & 3 & 0.3386 & 0.3359 & 0.3287 & 0.5408 & {\bf 0.2687} & 0.3596 \tn
& & (0.0040) & (0.0040) & (0.0032) & (0.1143) & (0.0557) & (0.0097) \tn
 & 7 & 2.1926 & 2.0864 & 2.0261 & 3.0027 & {\bf 1.4277} & 2.1116 \tn
& & (0.0338) & (0.0208) & (0.0155) & (0.5633) & (0.4497) & (0.0539) \tn
\hline
3-1 & 3 & 0.4944 & 0.4937 & 0.4907 & 0.5675 & {\bf 0.1457} & 0.4790 \tn
& & (0.0023) & (0.0024) & (0.0020) & (0.0109) & (0.0034) & (0.0013) \tn
 & 7 & 0.5555 & 0.5547 & 0.5427 & 0.7533 & {\bf 0.1866} & 0.4825 \tn
& & (0.0065) & (0.0059) & (0.0048) & (0.0184) & (0.0051) & (0.0016) \tn
\hline
3-2 & 3 & 0.1115 & 0.1102 & 0.1091 & 0.1191 & {\bf 0.0530} & 0.1083 \tn
& & (0.0014) & (0.0014) & (0.0011) & (0.0024) & (0.0017) & (0.0010) \tn
 & 7 & 0.1232 & 0.1214 & 0.1158 & 0.1368 & {\bf 0.0755} & 0.1144 \tn
& & (0.0025) & (0.0023) & (0.0017) & (0.0038) & (0.0028) & (0.0016) \tn
\hline
\end{tabular}
}
\caption{AMSE values over 100 simulated datasets with standard errors in parentheses when $n=300$. }
\label{table:MSEn300}
\end{table}

\clearpage

\begin{figure}
    \centering\includegraphics[scale=0.5]{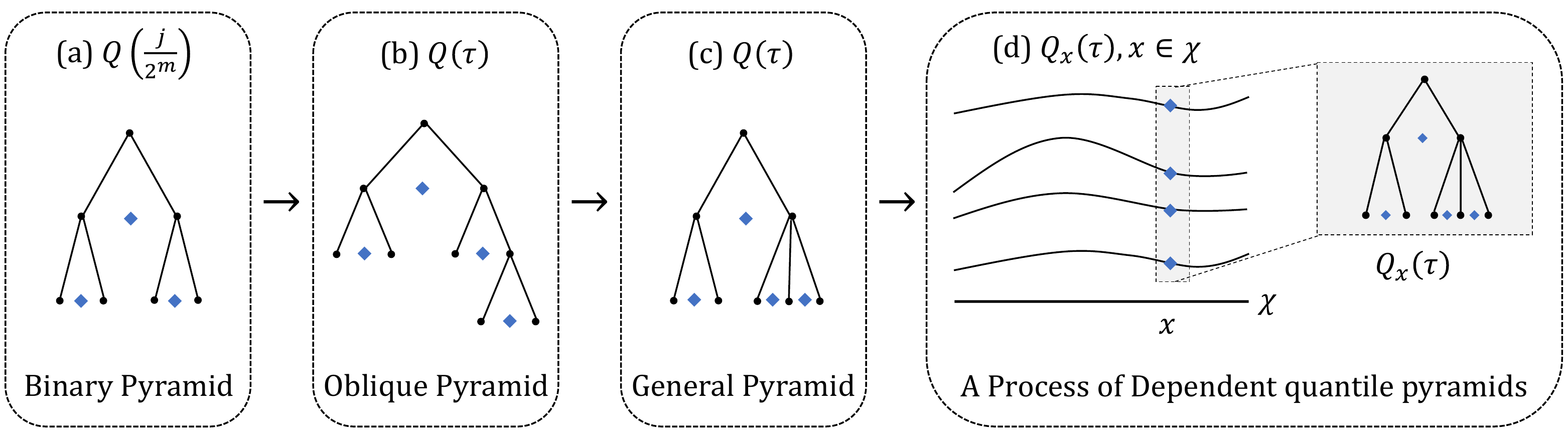}
    \caption{\small {\hy (a) A binary QP following \cite{hjort2009quantile} with three quartiles; (b) An oblique QP with four quantiles following \cite{rodrigues2019pyramid}; (c) A general QP with four quantiles; (d) A process of DQPs, where there is a QP at each value of $x$.} Each node in a tree structure corresponds to a subinterval. The first node is the unit interval. The rhombi represent quantiles pinned down in each subinterval.}
    \label{fig:DQPidea}
\end{figure}

\clearpage
\begin{figure}
    \centering\includegraphics[scale=0.42]{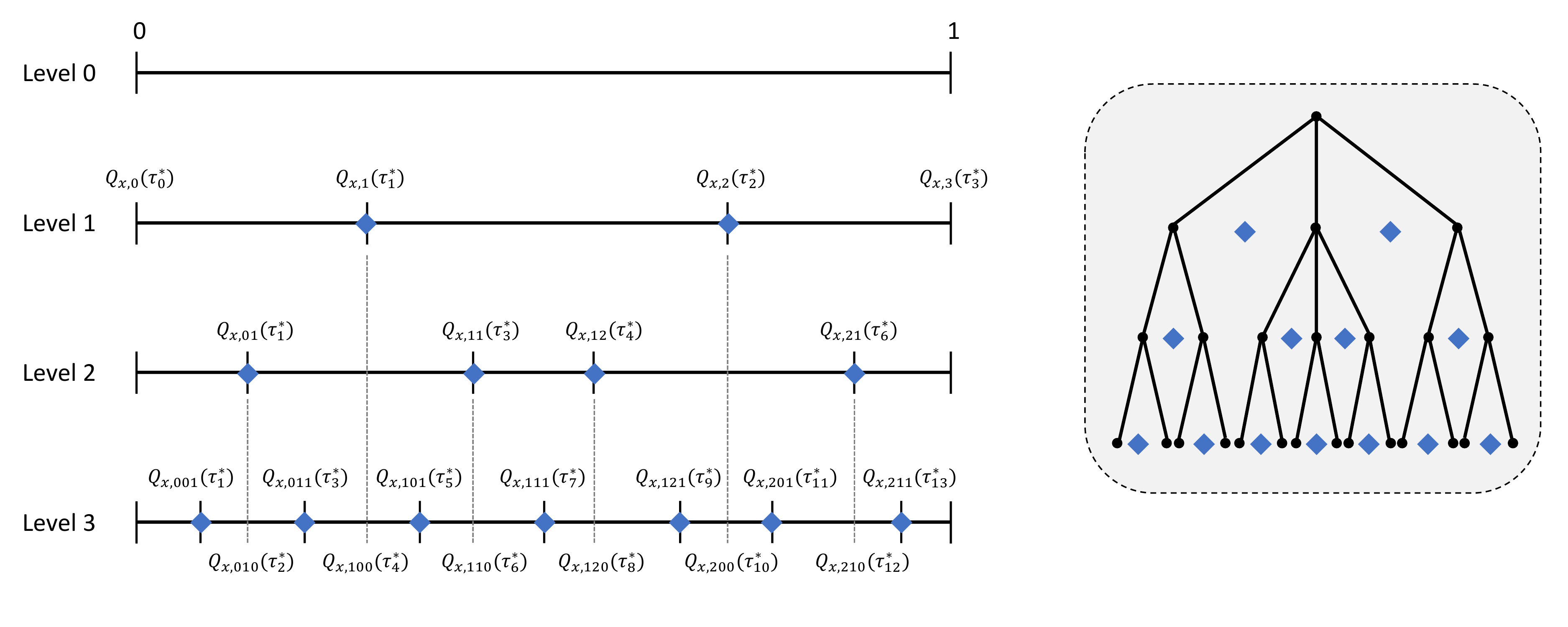}
    \caption{\small {\hy An example of a DQP with covariate $x$ with three levels and $13$ quantiles.  Both $\tau^*$ and $\epsilon$ notation is shown.  At level $0$, the interval at each $x$ is $[0,1]$.  At level $1$, $K=2$ quantiles are specified, creating three subintervals.  At level $2$, four quantiles are newly specified, leading to a total of $7$ subintervals.  Notation for the newly specified quantiles is shown.  At level $3$, one quantile is specified within each subinterval.  Both notations are shown for all quantiles in $(0,1)$.  Collecting the quantiles across $x$ gives the QR curves.  The corresponding tree structure is the same across all values of $x$.  It is shown in the right panel.  Dots and rhombi represent subintervals and quantiles, respectively.    
}
     }
    \label{fig:notation_DQP}
\end{figure}

\clearpage
\begin{figure}
    \centering\includegraphics[scale=0.6]{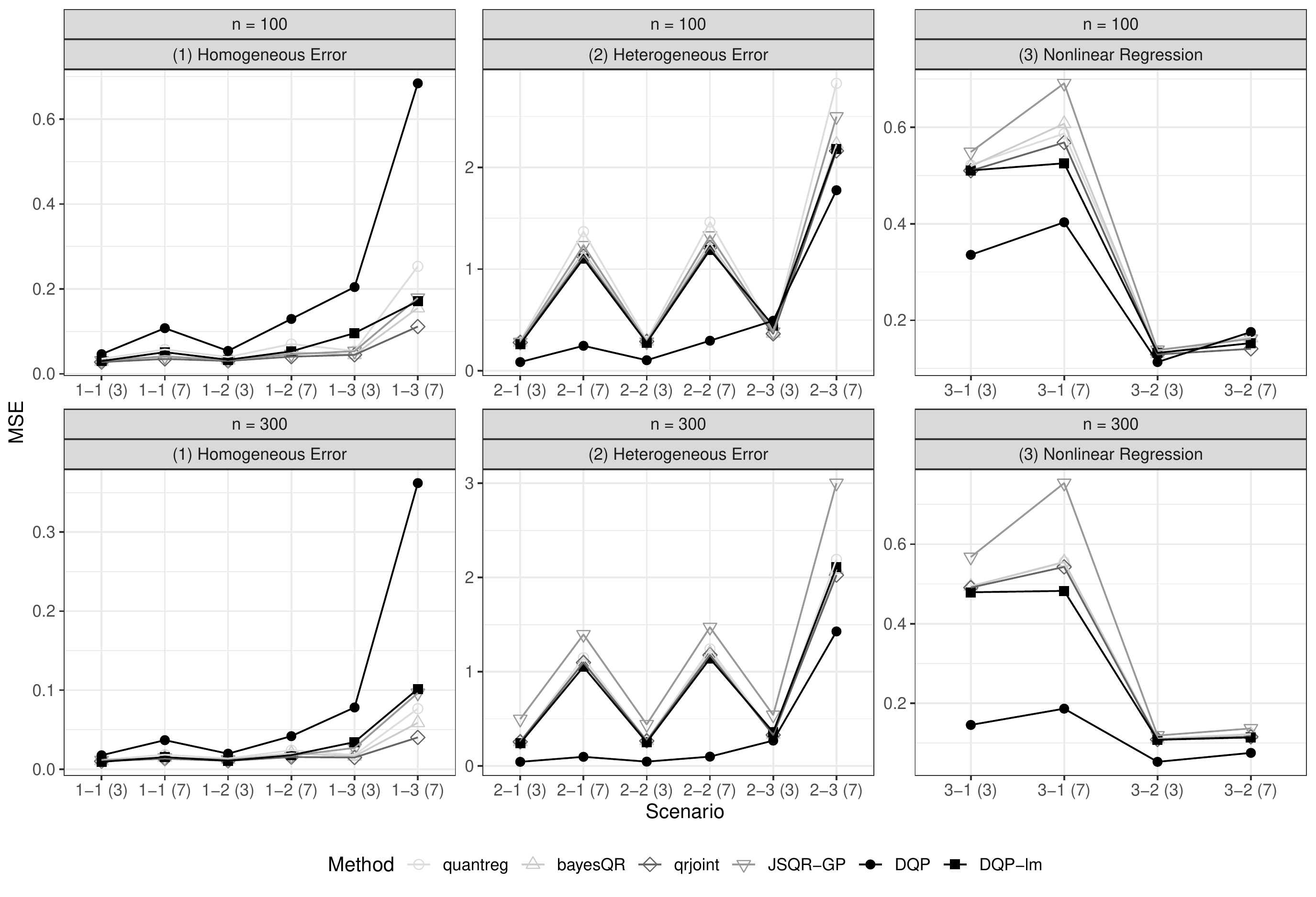}
    \caption{\small AMSE values for each combination of scenario-$T$-sample size}
    \label{fig:AMSE}
\end{figure}

\clearpage

\clearpage

\begin{figure}
\centering
\includegraphics[width=0.8 \linewidth]{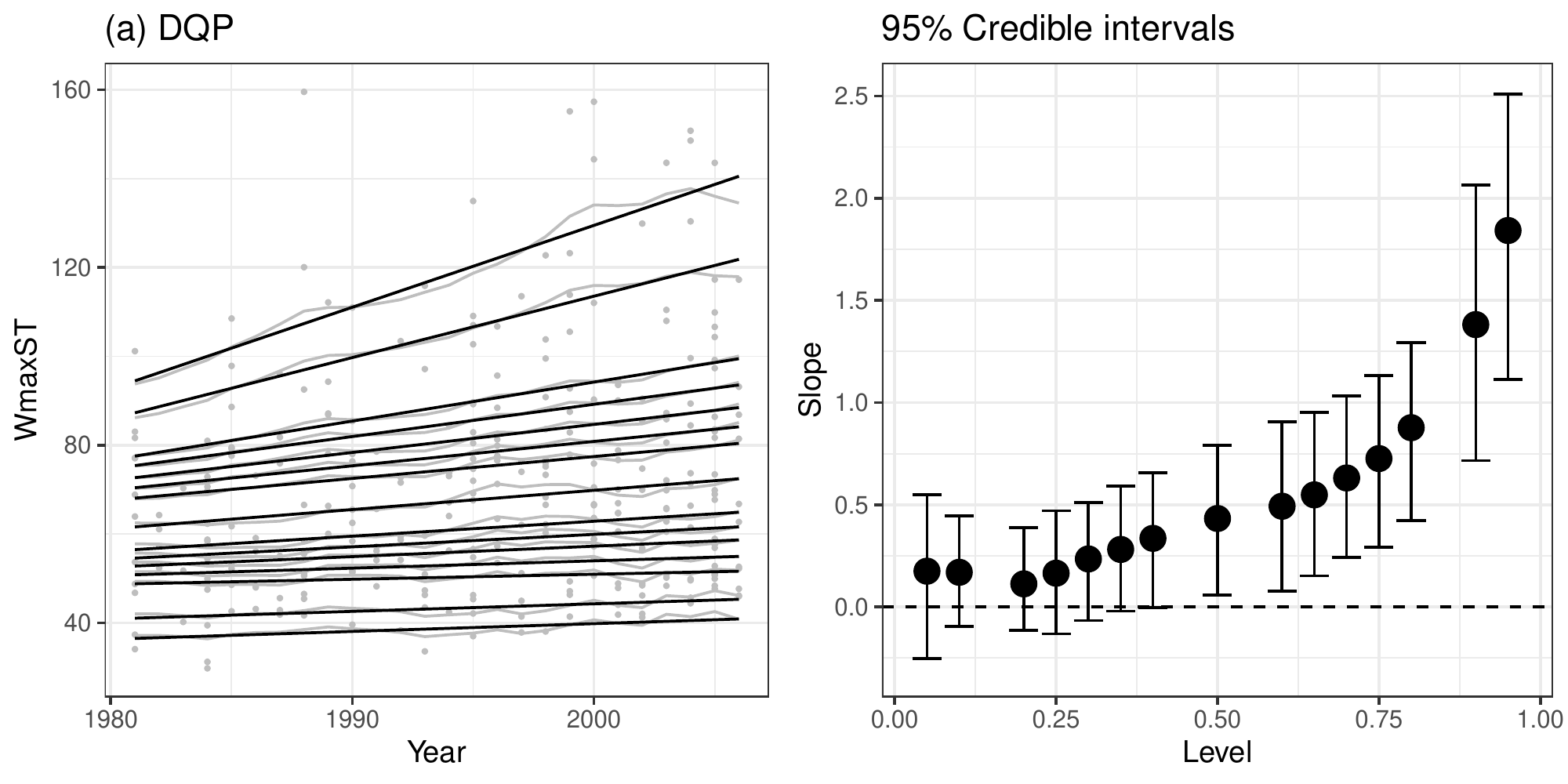}
\includegraphics[width=0.8 \linewidth]{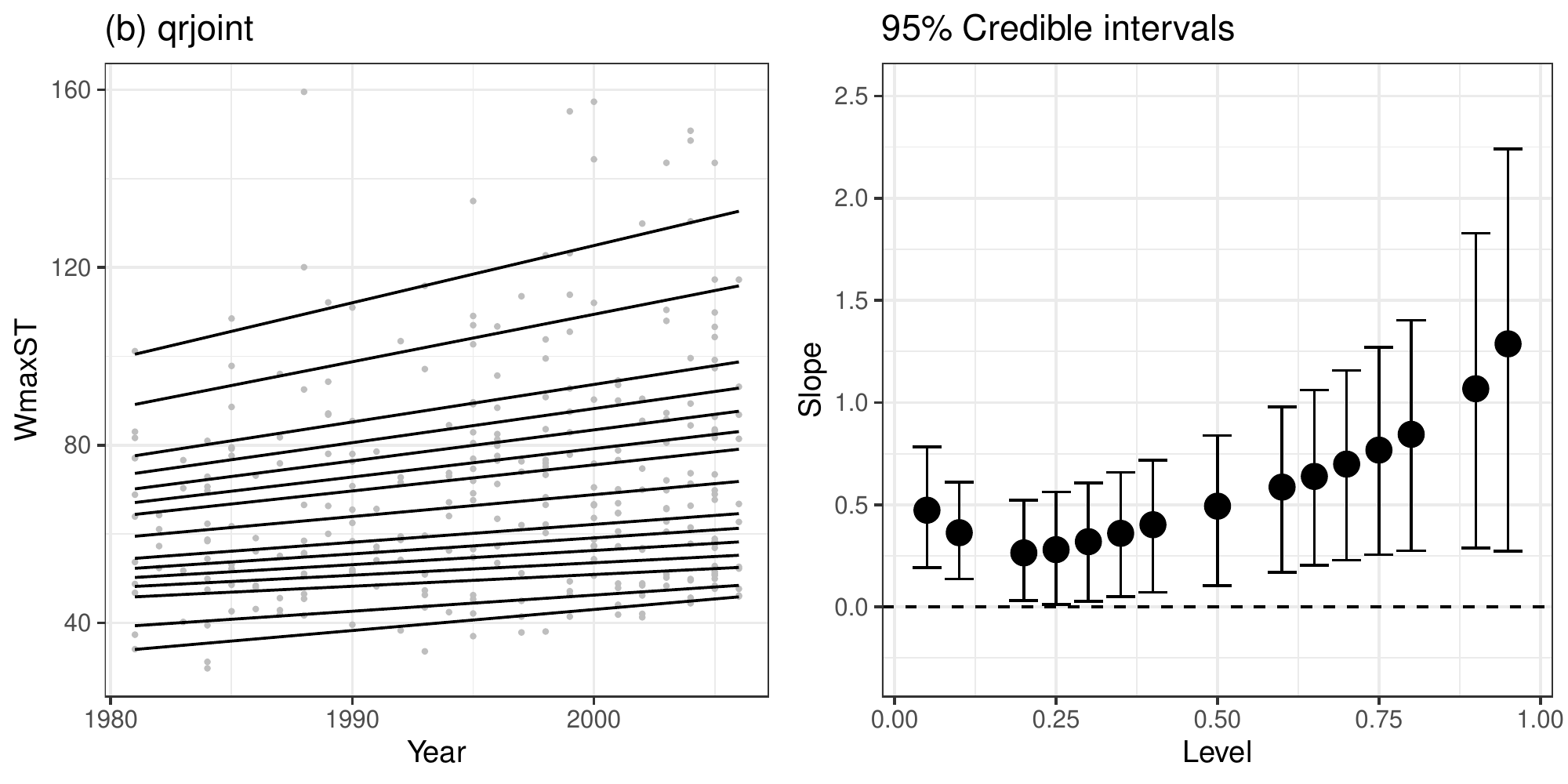}
\includegraphics[width=0.8 \linewidth]{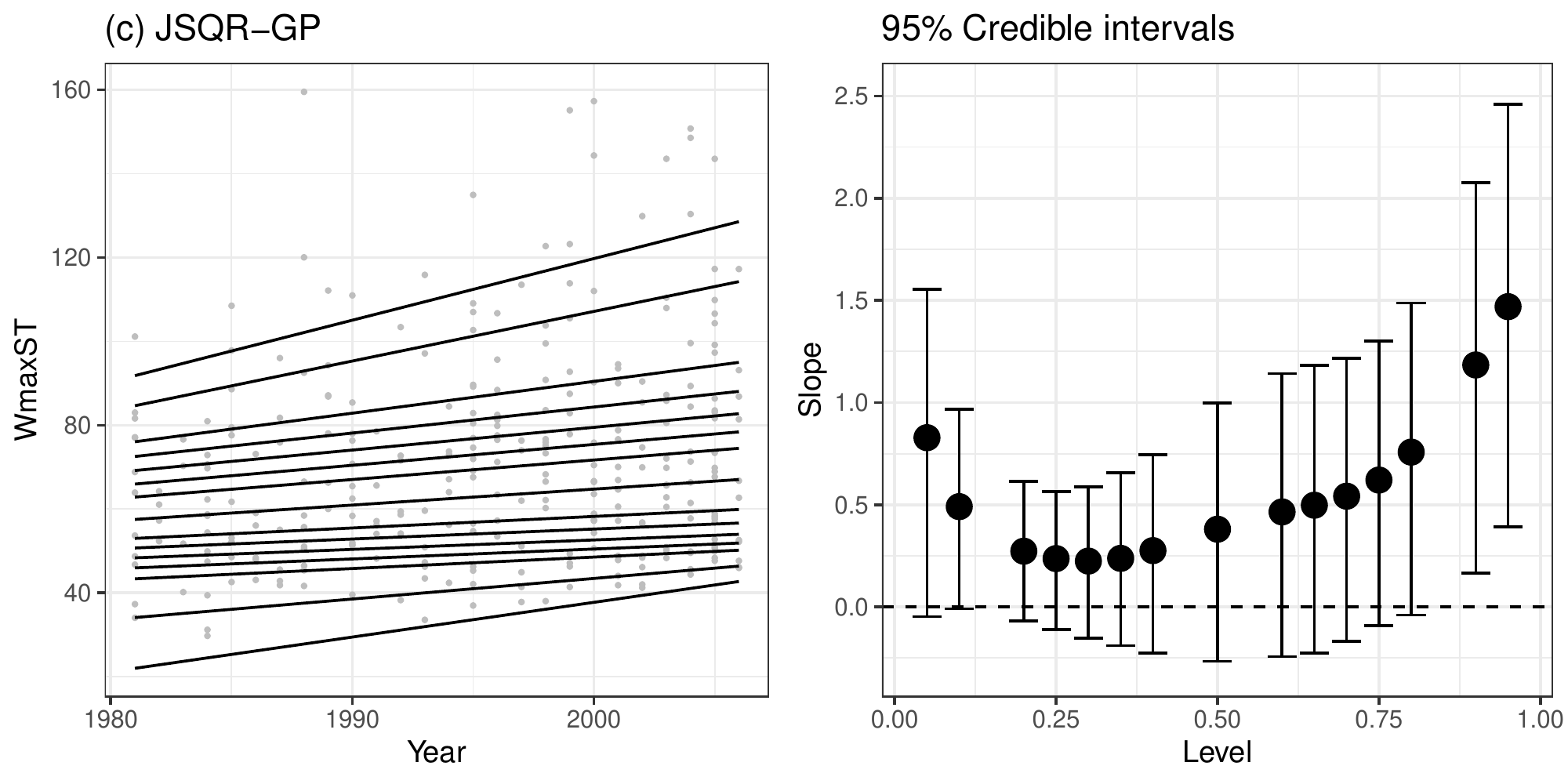}
	  \caption{\small \hy Left column: 15 estimated quantiles of cyclone intensity with grey data points. Right column: Estimated slopes of the quantile lines with 95\% empirical credible intervals. In (a), grey lines are the posterior means of the quantiles and black lines are the linearized fit of the posterior mean. }
	  	  \label{fig:cyclone_plots2}
\end{figure}


\clearpage
\section*{Figure captions}
\begin{itemize}
\item Figure~\ref{fig:DQPidea}: {\hy (a) A binary QP by \cite{hjort2009quantile} with three quartiles; (b) An oblique QP with four quantiles by \cite{rodrigues2019pyramid}; (c) A general QP with four quantiles; (d) A process of DQPs, where each QP corresponds to a value of $x$.} Each node in a tree structure corresponds to a subinterval. The first node is the unit interval. The rhombi represent quantiles pinned down in each subinterval.
\item Figure~\ref{fig:notation_DQP}: {\hy An example of a DQP with covariate $x$ with three levels and $13$ quantiles.  Both $\tau^*$ and $\epsilon$ notation is shown.  At level $0$, the interval at each $x$ is $[0,1]$.  At level $1$, $K=2$ quantiles are specified, creating three subintervals.  At level $2$, four quantiles are newly specified, leading to a total of $7$ subintervals.  Notation for the newly specified quantiles is shown.  At level $3$, one quantile is specified within each subinterval.  Both notations are shown for all quantiles in $(0,1)$.  Collecting the quantiles across $x$ gives the QR curves.  The corresponding tree structure is the same across all values of $x$.  It is shown in the right panel.  Dots and rhombi represent subintervals and quantiles, respectively.    
}
\item Figure~\ref{fig:AMSE}: AMSE values for each combination of scenario-$T$-sample size
\item Figure~\ref{fig:cyclone_plots2}: {\hy Left column: 15 estimated quantiles of cyclone intensity with grey data points. Right column: Estimated slopes of the quantile lines with 95\% empirical credible intervals. In (e), grey lines are the posterior means of the quantiles and black lines are the linearized fit of the posterior mean. }
\end{itemize}

\end{document}